\documentclass[twocolumn]{aastex62}

\usepackage{amssymb}
\usepackage{amsmath}
\usepackage{hyperref}
\usepackage{natbib}
\usepackage{color}
\usepackage{graphicx}% Include figure files
\usepackage{amsmath}
\usepackage{bm}

\usepackage{bm}

\accepted{}

\shorttitle{Electromagnetic Proton Beam Instabilities in the Inner Heliosphere}
\shortauthors{Liu, Zhao, et al.}
\begin{document}

\title{Electromagnetic Proton Beam Instabilities in the Inner Heliosphere: Energy Transfer Rate, Radial Distribution, and Effective Excitation}

\correspondingauthor{Jinsong Zhao}
\email{js\_zhao@pmo.ac.cn}

\author{Wen Liu}
\affiliation{Key Laboratory of Planetary Sciences, Purple Mountain Observatory, Chinese Academy of Sciences, Nanjing 210023,  People's Republic of China}
\affil{School of Astronomy and Space Science, University of Science and Technology of China, Hefei 230026, People's Republic of China}

\author{Jinsong Zhao}
\affiliation{Key Laboratory of Planetary Sciences, Purple Mountain Observatory, Chinese Academy of Sciences, Nanjing 210023,  People's Republic of China}
\affil{School of Astronomy and Space Science, University of Science and Technology of China, Hefei 230026, People's Republic of China}
\affil{Key Laboratory of Solar Activity, National Astronomical
Observatories, Chinese Academy of Sciences, Beijing 100012, China}

\author{Huasheng Xie}
\affiliation{Hebei Key Laboratory of Compact Fusion, Langfang 065001, People's Republic of China}
\affiliation{ENN Science and Technology Development Co., Ltd., Langfang 065001, People's Republic of China}

\author{Yuhang Yao}
\affiliation{Key Laboratory of Planetary Sciences, Purple Mountain Observatory, Chinese Academy of Sciences, Nanjing 210023,  People's Republic of China}
\affil{School of Astronomy and Space Science, University of Science and Technology of China, Hefei 230026, People's Republic of China}

\author{Dejin Wu}
\affiliation{Key Laboratory of Planetary Sciences, Purple Mountain Observatory, Chinese Academy of Sciences, Nanjing 210023,  People's Republic of China}

\author{L. C. Lee}
\affiliation{Institute of Earth Sciences, Academia Sinica, Taipei, 11529, Taiwan}
\affil{Space Science Institute, Macau University of Science and Technology, Macau, People's Republic of China}

\begin{abstract}

Differential flows among different ion species are often observed in the solar wind, and such ion differential flows can provide the free energy to drive Alfv\'en/ion-cyclotron and fast-magnetosonic/whistler instabilities. Previous works mainly focused on ion beam instability under the parameters representative of the solar wind nearby 1 au. In this paper we further study proton beam instability using the radial models of the magnetic field and plasma parameters in the inner heliosphere. We explore a comprehensive distribution of proton beam instability as functions of the heliocentric distance and the beam speed. We also perform a detailed analysis of the energy transfer between unstable waves and particles and quantify how much the free energy of the proton beam flows into unstable waves and other kinds of particle species (i.e., proton core, alpha particle, and electron). 
This work clarifies that both parallel and perpendicular electric fields are responsible for the excitation of oblique Alfv\'en/ion-cyclotron and oblique fast-magnetosonic/whistler instabilities. Moreover, this work proposes an effective growth length to estimate whether the instability is efficiently excited or not. It shows that oblique Alfv\'en/ion-cyclotron instability, oblique fast-magnetosonic/whistler instability, and oblique Alfv\'en/ion-beam instability can be efficiently driven by proton beams drifting at the speed $\sim 600-1300$ km s$^{-1}$ in the solar atmosphere. In particular, oblique Alfv\'en/ion-cyclotron waves driven  in the solar atmosphere can be significantly damped therein, leading to the solar corona heating. These results are helpful for understanding proton beam dynamics in the inner heliosphere and can be verified through in situ satellite measurements.

\end{abstract}


\section{Introduction} \label{sec:introduction}

The proton velocity distribution in the solar wind usually consists of two components \citep[]{1973JGR....78.2017F,1974RvGSP..12..715F,1982JGR....87...52M}: a more dense core and a secondary tenuous beam that drifts at a speed faster than the core. The proton beam component was firstly found by \cite{1973JGR....78.2017F} from the IMP spacecrafts. Helios observations further identified the proton beam population arising from heliocentric distance 0.3 $-$1 au \citep{1982JGR....87...52M}. The statistical analysis for the Helios and Ulysses data sets have explored that the differential drift speed $\Delta V$ between the core and beam components is of the order of the local Alfv\'en speed $V_A$ \citep[]{1987JGR....92.7263M,2000GeoRL..27...53G,2004JGRA..109.5101T,2018ApJ...864..112A,2019SoPh..294...97D}. The observed $\Delta V/V_A$ are normally less than the values predicted from proton beam instability \citep[]{1987JGR....92.7263M,2000GeoRL..27...53G}. Consequently, proton beam instability is proposed to play a significant role in constraining the proton beam in the solar wind  \citep[]{1987JGR....92.7263M,2000GeoRL..27...53G}.

The proton beam provides one of free energies to drive the electromagnetic instabilities. \cite{1975PhRvL..35..667M,1976JGR....81.2743M} firstly performed a comprehensive investigation of proton beam instability in a plasma containing proton core, proton beam,  and electron components, and found that the proton beam induces three kinds of instabilities: oblique Alfv\'en/ion-cyclotron instability, oblique fast-magnetosonic/whistler instability, and parallel fast-magnetosonic/whistler instability. \cite{1998JGR...10320613D} reconsidered proton beam instability, and identified the appearance of two kinds of oblique Alfv\'en/ion-cyclotron instabilities, i.e., Alfv\'en I occurring at comparatively short wavelengths \citep[also see][]{1992JGR....9714779W}, and Alfv\'en II at comparatively longer wavelengths \citep[also see][]{1975PhRvL..35..667M,1976JGR....81.2743M}. Besides, the proton beam could drive parallel Alfv\'en/ion-cyclotron instability that have a speed threshold higher than that for parallel fast-magnetosonic/whistler instability \citep[e.g.,][]{1991SSRv...56..373G,2019ApJ...874..128L}. Among these instabilities, parallel fast-magnetosonic/whistler and Alfv\'en I instabilities are the two strongest instabilities under plasma parameters representative of 1 au solar wind, and they are thought of as the candidates constraining the proton beam therein \citep[]{1998JGR...10320613D,1999JGR...104.4657D}.

In contrast to studies of proton beam instability in the vicinity of the solar wind at 1 au, there are a few works that have focused on such instability in the region close to the Sun. Recently, \cite{2019ApJ...874..128L} proposed that parallel Alfv\'en/ion-cyclotron and fast-magnetosonic/whistler instabilities can be driven by the proton beam in the solar coronal holes. Using the Parker Solar Probe (PSP) measurements, \cite{2020ApJS..248....5V} reported observations of the simultaneous occurrence of proton beams and ion-scale waves at heliocentric distances of about $36R_S$, and identified that the observed waves are locally driven by proton beams on the basis of the instability analysis. \cite{2020ApJS..246...66B} showed that ion-scale waves are observed $30\% - 50\%$ radial field intervals in the first encounter of PSP. Moreover, in comparison to the proton beam speed comparable to $V_A$ in the solar wind nearby 1 au, proton beams are at times seen by PSP with relative speeds $\gtrsim1.5V_A$ \citep{{2021ApJ...909....7K}}.
Since PSP will measure the ion velocity distribution and electromagnetic fields down to the heliocentric distance at $9.8R_s$, it provides a unique opportunity to identify the excitation of proton beam instability in the solar atmosphere. Consequently, this paper plans to investigate proton beam instability in both the solar atmosphere and the solar wind. 

In this paper, different from previous works that use parameters representative of one location in the solar wind, we study the ion-scale proton beam instability under parameters radially distributed in the heliocentric distance from $3-215R_S$. This study explores the nature of four typical instabilities, i.e., oblique Alfv\'en/ion-cyclotron instability, oblique fast-magnetosonic/whistler instability, oblique Alfv\'en/ion-beam instability, and parallel fast-magnetosonic/whistler instability, in the inner heliosphere. In particular, this study recognizes that the Alfv\'en I instability proposed by \cite{1998JGR...10320613D} comes from the coupling between the Alfv\'en/alpha-cyclotron mode and the Alfv\'en/ion-beam mode. Also, this study explores the excitation mechanism of each kind of instability using the energy transfer rate between unstable waves and particles in both the parallel and perpendicular directions with respect to the background magnetic field, which clearly shows how much the proton beam energy flows into unstable waves and other particle components in each kind of instability. Furthermore, this study presents the controlling parameter region of each proton beam instability and proposes that oblique Alfv\'en/ion-cyclotron, oblique fast-magnetosonic/whistler, and Alfv\'en/ion-beam instabilities could be driven by the proton beam with the drift speed $\sim 600-1300$ km s$^{-1}$ in the solar atmosphere, which can be checked by PSP observations. 

This paper is organized as follows. Section 2 introduces the theoretical model and plasma parameters. Section 3 analyzes the nature and excitation mechanism of four typical proton beam instabilities. Section 4 gives the radial distributions of proton beam instability in the inner heliosphere. Section 5 considers the effective excitation of the instability. Section 6 discusses the change in the plasma temperature during proton beam instability and shows the dependence of the instability on plasma parameters, such as the temperature anisotropy, the differential drift of alpha particles relative to protons, and the relative proton beam density. Lastly, our results are summarized in Section 7.

\section{Theoretical Model and Plasma Parameters}

\subsection{Theoretical Model}
\label{sub:2.1}
To study the wave dynamics in the weakly collisional solar wind plasma, we use the model consisting of Vlasov and Maxwell's equations, which yield the wave equation in Fourier space 

\begin{equation}
\mathbf{k} \times \left(\mathbf{k}\times \mathbf{E} \right) + \frac{\omega^2}{c^2} {\bm \epsilon} \cdot \mathbf{E},
\label{dispersion equation}
\end{equation}
where ${\bm \epsilon} = i {\bm \sigma} /({\epsilon_0\omega})+ \mathbf{I}$, $\epsilon_0$ is the permittivity of free space, ${\bm \epsilon}$ is the dielectric tensor, ${\bm \sigma}$ is the conductivity tensor, $\omega$ is the wave frequency, and $\mathbf{E}$ is the wave electric field. The plasma wave eigenmodes correspond to solutions of Equation (\ref{dispersion equation}). Recently, a general dispersion relation solver named BO/PDRK for Equation (\ref{dispersion equation}) was developed by \cite{2016PlST...18...97X} and \cite{2019CoPhC.244..343X}, and this solver is useful performing a comprehensive study for ion and electron kinetic instabilities \citep[]{2019ApJ...884...44S,2020ApJ...902...59S}. In this paper we use BO/PDRK to give the wave dispersion relation in proton beam plasma. 

One key problem in the kinetic instability study is the role of Landau and cyclotron resonances between unstable waves and particles on the instability excitation. Both Landau and cyclotron resonances can induce the free energy of particles flowing into plasma waves, resulting in wave amplification \citep[e.g.,][]{1991SSRv...56..373G}. To estimate the contribution of Landau and cyclotron resonances, a popular method is to calculate the resonance factor $\eta_{\mathrm{sn}} = \left( \omega - k_\parallel V_{s} - n \Omega_{\mathrm{cs}} \right) / \sqrt{2}k_\parallel V_{\mathrm{Ts}}$ \citep[e.g.,][]{1991SSRv...56..373G}, where $V_{s}$ is the drift velocity along the background magnetic field $\mathbf{B_0}$, $V_{\mathrm{Ts}}\equiv \sqrt{T_s/m_s}$ is the thermal speed, $\Omega_{\mathrm{cs}}\equiv q_sB_0/m_s$ is the cyclotron frequency, and ``s'' denotes the particle species. Normally, $|\eta_{\mathrm{s0}}|\lesssim 2$ (or $3$) indicates that the Landau resonance interaction is important in triggering instability; and $|\eta_{\mathrm{sn}}(n\neq0)| \lesssim 2$ (or $3$) implies that the cyclotron resonance interaction is in favor of triggering instability. We note that the transit-time resonant interaction can also arise as $\eta_{\mathrm{s0}}\sim 0$  \citep[e.g.,][]{1992wapl.book.....S,1998ApJ...500..978Q}. 

This study will use an alternative parameter, i.e., the energy transfer rate, to quantify wave-particle resonances on the instability excitation. We will calculate the energy transfer rate by using the plasma current $\mathbf{J}$ and the wave electric field $\mathbf{E}$. The similar energy absorption/emission calculation has been proposed by previous works \citep[e.g.,][]{1992wapl.book.....S,1998ApJ...500..978Q,2017JPlPh..83a7002H,2017JPlPh..83d5301K,2019ApJ...887..234K,2020JPlPh..86d9002K,2020ApJ...898...43H}. Once the wave frequency and wave electric field fluctuations are obtained from Equation (\ref{dispersion equation}), the plasma current is given by

\begin{equation}
\mathbf{J_s} = {\bm \sigma_s} \cdot \mathbf{E}.
\label{eq:J}
\end{equation}

The energy transfer rate between the waves and particles can be quantified by
\begin{equation}
\Gamma_s = \frac{1}{4}\left( \mathbf{E} \cdot\mathbf{J^*_s} + \mathbf{E}^* \cdot \mathbf{J_s}  \right),
\end{equation}
which denotes the energy absorption/emission per unit of time, and per unit of volume. $\Gamma_s$ can be further decomposed as contributions from the parallel and perpendicular electric fields:
\begin{equation}
\Gamma_{\mathrm{s\parallel}} = \frac{1}{4}\left( \mathbf{E}_\parallel \cdot \mathbf{J^*_{s\parallel}} + \mathbf{E^*_\parallel} \cdot \mathbf{J_{s\parallel}} \right),
\end{equation}
and 
\begin{equation}
\Gamma_{\mathrm{s\perp}} = \frac{1}{4}\left( \mathbf{E}_\perp \cdot \mathbf{J^*_{s\perp}} + \mathbf{E^*_\perp} \cdot \mathbf{J_{s\perp}} \right).
\end{equation}
The total energy transfer rate can be obtained by summing all particle energy transfer rates: $\Gamma_t = \sum_s \Gamma_s$, $\Gamma_{\mathrm{t\parallel}}= \sum_s \Gamma_{\mathrm{s\parallel}}$, and $\Gamma_{\mathrm{t\perp}}= \sum_s \Gamma_{\mathrm{s\perp}}$.
Since Landau and cyclotron resonances are dependent on parallel and perpendicular electric field fluctuations, respectively, they can be quantitatively measured by using $\Gamma_{\parallel}$ and $\Gamma_{\perp}$. When the energy transfer rate is smaller than zero, the energy transfers from particles into waves, leading to wave growth. While the energy transfer rate is larger than zero, the energy transfers from waves into particles, leading to wave damping. 

In this study we use the following expressions to quantify the energy transfer between the waves and particles,
\begin{eqnarray}
P_s = \frac{ \Gamma_s }{W_{\mathrm{EB}}}, ~
P_{\mathrm{s\parallel}} = \frac{\Gamma_{\mathrm{s\parallel}}}{W_{\mathrm{EB}}},~
P_{\mathrm{s\perp}} = \frac{ \Gamma_{\mathrm{s\perp}} }{W_{\mathrm{EB}}},
\label{Ps_1}
\end{eqnarray}
and
\begin{equation}
P_t = \sum_s P_s, ~ P_{\mathrm{t\parallel}}= \sum_s P_{\mathrm{s\parallel}}, ~P_{\mathrm{t\perp}}= \sum_s P_{\mathrm{s\perp}}.
\label{Ps_2}
\end{equation}
where $W_{\mathrm{EB}}=\epsilon_0|\mathbf{E}|^2/4 + |\mathbf{B}|^2/4\mu_0$ is the wave electromagnetic energy. These expressions quantify the energy absorption/emission per unit of time, per unit of volume, and per unit of wave electromagnetic energy. We note that a different definition for the energy transfer is proposed in previous studies \citep[e.g.,][]{1998ApJ...500..978Q}, and the relation between these two expressions is discussed in Appendix A.
One advantage of using these normalized rates is that due to $P_t/2=-\gamma$ (see Appendix A), we can directly measure the contribution of each resonance effect on wave growth or damping. Here, $\gamma$ represents the imaginary part of $\omega$ in which $\gamma>0$ (or $<0$) corresponds to wave growth (or damping). In addition, we use $\omega$ to calculate $P_t$, not the real part of $\omega$, and the reason is that a complete plasma current is only obtained by using $\omega$ (see Equation \ref{eq:J}).

Furthermore, we will decompose the energy transfer rate at different $n$, i.e., $P_s(n)$, and investigate the contribution of $n=0$ and $n\neq 0$ resonances on wave growth or damping. Based on these energy transfer rates, we will give detailed analyses for the growth and damping mechanism in each proton beam instability in Section 3.

\subsection{Magnetic Field and Plasma Parameters}
\label{sub:2.2}

\begin{figure}[h]
\centerline{
  \includegraphics[width=\columnwidth]{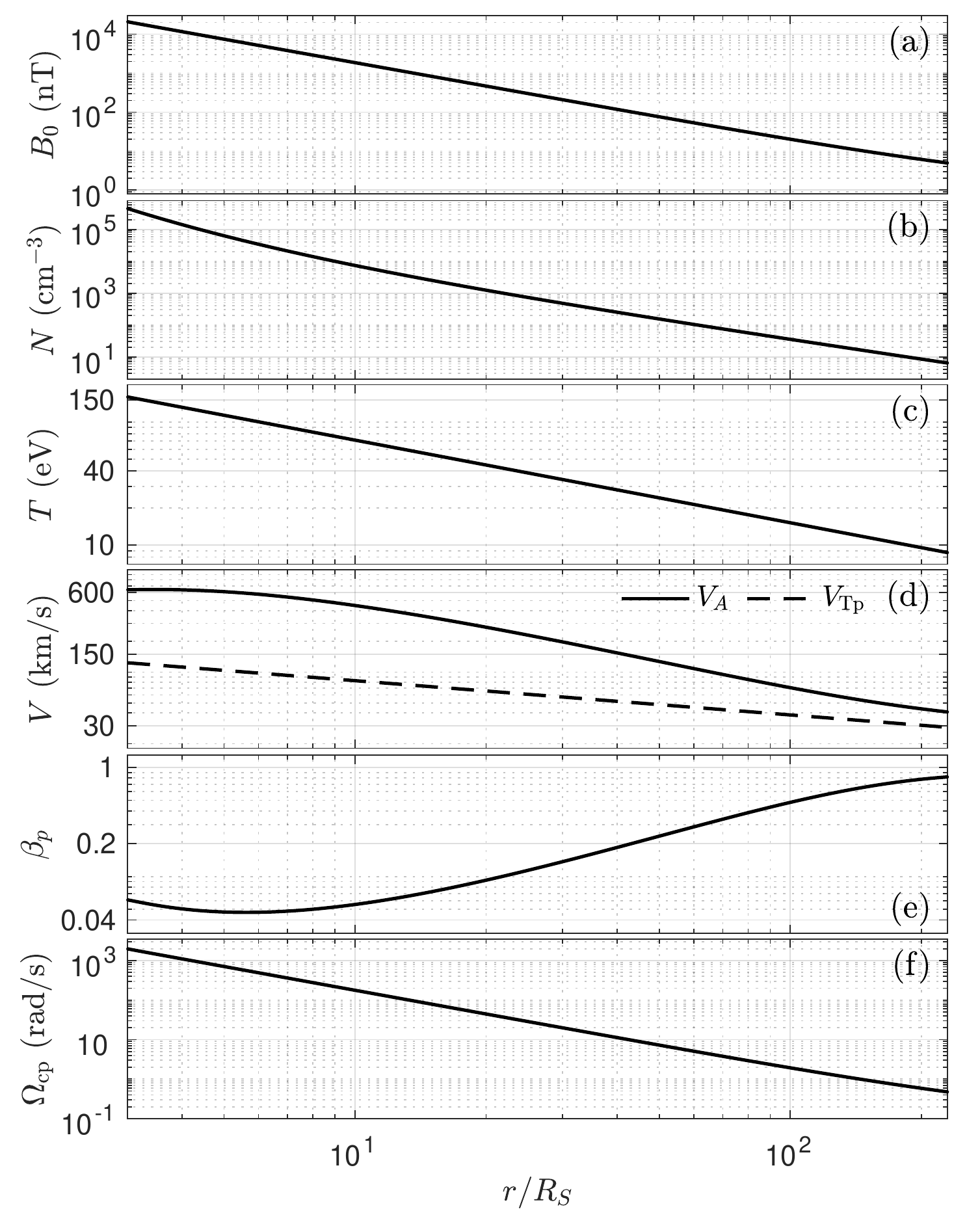}}
\caption{
The radial distributions of the magnetic field strength and plasma parameters. (a) The magnetic field strength $B_0$; (b) the plasma number density $N$; (c) the plasma temperature $T$; (d) the Alfv\'en speed $V_A$ (solid line), and proton thermal speed $V_{\mathrm{Tp}}$ (dashed line); (e) the plasma proton beta $\beta_p$ defined as the ratio of the proton thermal pressure to magnetic pressure; and (f) the proton cyclotron frequency $\Omega_{\mathrm{cp}}$.
\label{fig:parameters}}
\end{figure}

PSP will measure plasma waves and particle velocity distributions in situ down to locations below $10R_S$ in the inner heliosphere.  
In order to provide a direct comparison between PSP observations and theoretical instability predictions, we use radial distributions of the magnetic field strength and plasma parameters at PSP orbits referring to \cite{2016SSRv..204...49B}. The detailed fitting procedures are stated in \cite{2016SSRv..204...49B}, and here we merely give the results.

The magnetic field strength is
\begin{equation}
B_0 = \frac{860R_S}{r}\sqrt{\left(\frac{215R_S}{r} \right)^2 + \left(\frac{405}{V_{\mathrm{sw}}} \right)^2} ~(\mathrm{nT}),
\end{equation}
where $r$ is the heliocentric distance, $R_S$ is the solar radius, and the solar wind velocity is given as $V_{\mathrm{sw}} = 430 \sqrt{1- \mathrm{exp} \left (-\frac{r/R_S-2.8}{25} \right)} ~(\mathrm{km/s}).$
The electron number density is \citep[also see][]{1999ApJ...523..812S}
\begin{equation}
N_e =  N_{0}\times \mathrm{exp}\left( \frac{3.67R_S}{r}  \right)\left( \frac{R_S^2}{r^2} + \frac{4.9R_S^3}{r^3} + \frac{7.6R_S^4}{r^4} +  \frac{6.0R_S^5}{r^5} \right)
\end{equation}
with $N_{0}=3.26\times10^5$ cm$^{-3}$, and the proton temperature is given by
\begin{equation}
T_p = \frac{T_{0}}{(r/R_S)^{0.6}},
\end{equation}
where $T_{0}=226.4$ eV. 

We consider a plasma containing four particle components, i.e., electrons ``$e$'',  proton core ``$pc$'', proton beam ``$pb$'' and alpha particles ``$\alpha$'', and assume their velocity distribution functions following the drifting Maxwellian distribution, i.e., $f_s\left( v_\parallel,v_\perp \right) = \frac{N_s} {(\pi)^{3/2} (2T_s/m_s)^{3/2}} \mathrm{exp} \left[-\frac{\left(v_\parallel-V_s\right)^2 + v_\perp^2}{2T_s/m_s}  \right]$, where $m_s$, $N_s$, $T_s$ and $V_s$ denote the mass, number density, temperature, and drift speed for each particle component ``$s$''. 
We also consider the proton core frame, which is $V_{\mathrm{pc}}=0$. To better show the pure proton beam instability, we assume there is no differential drift between alpha particle and proton core components, i.e., $V_{\alpha}=0$, while in fact, alpha particles are streaming faster than core protons in the solar wind \citep[e.g.,][]{1982JGR....87...35M,2018ApJ...864..112A,2019SoPh..294...97D}. Moreover, we assume $N_{\mathrm{pb}}V_{\mathrm{pb}}-N_eV_e=0$ to ensure a zero current condition. For the number density and temperature of each particle component, we use following values: $N_{\mathrm{pc}}=0.8N_e$, $N_{\mathrm{pb}}=0.1N_e$, and $N_{\alpha}=0.05N_e$; and $T_{\mathrm{pc}}=T_{\mathrm{pb}}=T_{\alpha}=T_e$. The discussion of the dependence of the instability on the particle density and temperature will be given in Section 6.

The magnetic field strength and plasma parameters as a function of the heliocentric distance $r$ are presented in Figure ~\ref{fig:parameters}. Figure ~\ref{fig:parameters} also gives the radial distributions of the Alfv\'en speed $V_A$, the proton thermal speed $V_{\mathrm{Tp}}$, the plasma proton beta $\beta_p$ (the ratio of the proton thermal to magnetic pressure), and the proton cyclotron frequency $\Omega_{\mathrm{cp}}$. It should be emphasized that $\beta_p$ is one important parameter affecting proton beam instability \citep[]{1976JGR....81.2743M,1998JGR...10320613D}. From Figure ~\ref{fig:parameters}, we see that $\beta_p$ is smaller than 0.1 as $r\lesssim 20R_S$, and then $\beta_p$ increases with $r$ and is about $0.9$ at $r=215R_S$ (1 au). 

\subsection{Basic Wave Modes}
\label{sub:2.3}

\begin{figure}[h]
\centerline{
  \includegraphics[width=\columnwidth]{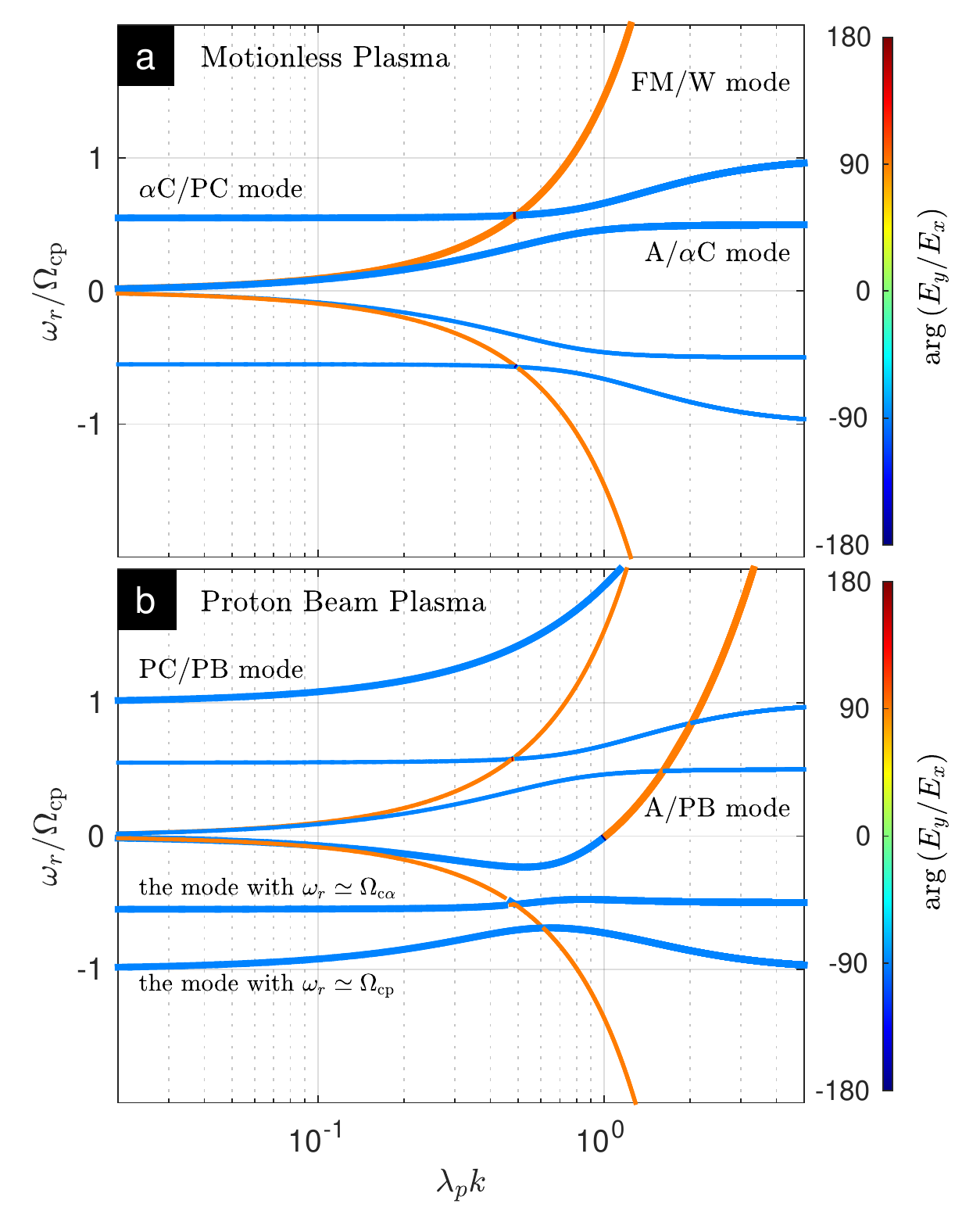}}
\caption{
Parallel and antiparallel plasma waves in (a) a proton-alpha-electron plasma and (b) a plasma containing proton core, proton beam, alpha particle, and electron components. The argument of $E_y/E_x$ is overlaid on the wave dispersion relations. 
Three wave modes correspond to the A/$\alpha$C (Alfv\'en/alpha-cyclotron) mode, the $\alpha$C/PC (alpha-cyclotron/proton-cyclotron) mode, and the FM/W (fast-magnetosonic/whistler) mode in panel (a); and in panel (b) the wave modes significantly affected by the proton beam are the PC/PB (proton-cyclotron/proton-beam) mode, the A/PB (Alfv\'en/proton-beam) mode, the mode with the frequency near $\Omega_{\mathrm{c\alpha}}$, and the mode with the frequency near $\Omega_{\mathrm{cp}}$.
\label{fig:WaveFluid}}
\end{figure}

For identifying which kind of the wave mode is unstable, this subsection introduces the basic wave modes in proton beam plasmas. The proton beam can considerably affect the dispersion relations of plasma waves \citep{2019ApJ...874..128L}, and an example is given in Figure~\ref{fig:WaveFluid}, which presents all parallel and antiparallel low-frequency waves in a cold plasma.
In an electron-proton-alpha particle plasma without any relative drifts between particle components, there are three kinds of wave modes below the electron cyclotron frequency: Alfv\'en/alpha-cyclotron mode, alpha-cyclotron/proton-cyclotron mode, and fast-magnetosonic/whistler mode (Figure~\ref{fig:WaveFluid}a). 
When the plasma contains a proton beam with a drift speed $V_{\mathrm{pb}}=V_A$ (Figure~\ref{fig:WaveFluid}b), the coupling between the backward Alfv\'en/alpha-cyclotron mode and the proton beam mode results in the appearance of an Alfv\'en/proton-beam mode. The short-wavelength Alfv\'en/proton-beam mode wave becomes right-hand polarization and forward propagation due to the effect of the Doppler shift frequency $V_{\mathrm{pb}}k_z$. Also, a new proton-cyclotron/proton-beam mode arises in the forward propagation direction, and two left-hand polarized wave modes in the backward direction correspond to a mode with frequency near the alpha cyclotron frequency and a mode with frequency near the proton cyclotron frequency.

For the proton beam propagating against the background magnetic field, the basic wave modes are the same as that in Figure~\ref{fig:WaveFluid}b, and only the difference is the wave direction. Therefore, this study considers the situation where the proton beam propagates along the background magnetic field.

\section{Four typical proton beam instabilities}

\begin{figure*}[!htb]
\includegraphics[width=\textwidth]{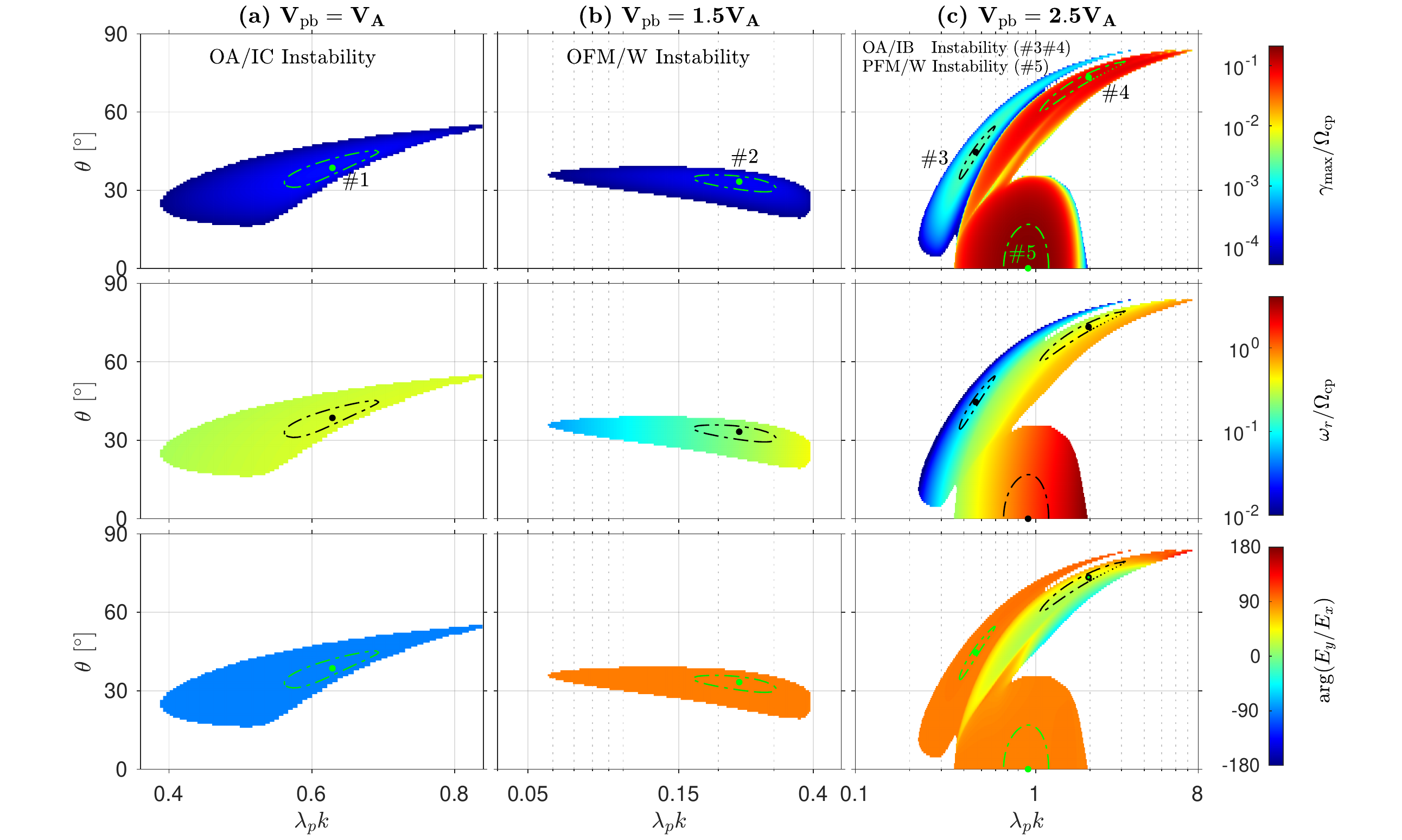}
\caption{
The $k-\theta$ distributions of four typical proton beam instabilities at $r=10R_S$: (a) the OA/IC instability driven by the $V_{\mathrm{pb}}=V_A$ proton beam; (b) the OFM/W instability driven by the $V_{\mathrm{pb}}=1.5V_A$ proton beam; and (c) the OA/IB instability and PFM/W instability driven by the $V_{\mathrm{pb}}=2.5V_A$ proton beam. (Top panels) the maximum growth rate, $\gamma_{\mathrm{max}}$; (middle panels) the real frequency $\omega_r$ at $\gamma_{\mathrm{max}}$; and (bottom panels) the argument of $E_y/E_x$ at $\gamma_{\mathrm{max}}$. The circular points marked by $\#1$, $\#2$, $\#3$, $\#4$ and $\#5$ denote local maximum growth rates, and the counter lines correspond to 0.9 times each local maximum growth rate. 
OA/IB = oblique Alfv\'en/ion-beam; OA/IC = oblique Alfv\'en/ion-cyclotron; OFM/W = oblique fast-magnetosonic/whistler; and PFM/W = parallel fast-magnetosonic/whistler.
\label{fig:k-theta_10Rs}}
\end{figure*}

\begin{figure*}
\includegraphics[width=\textwidth]{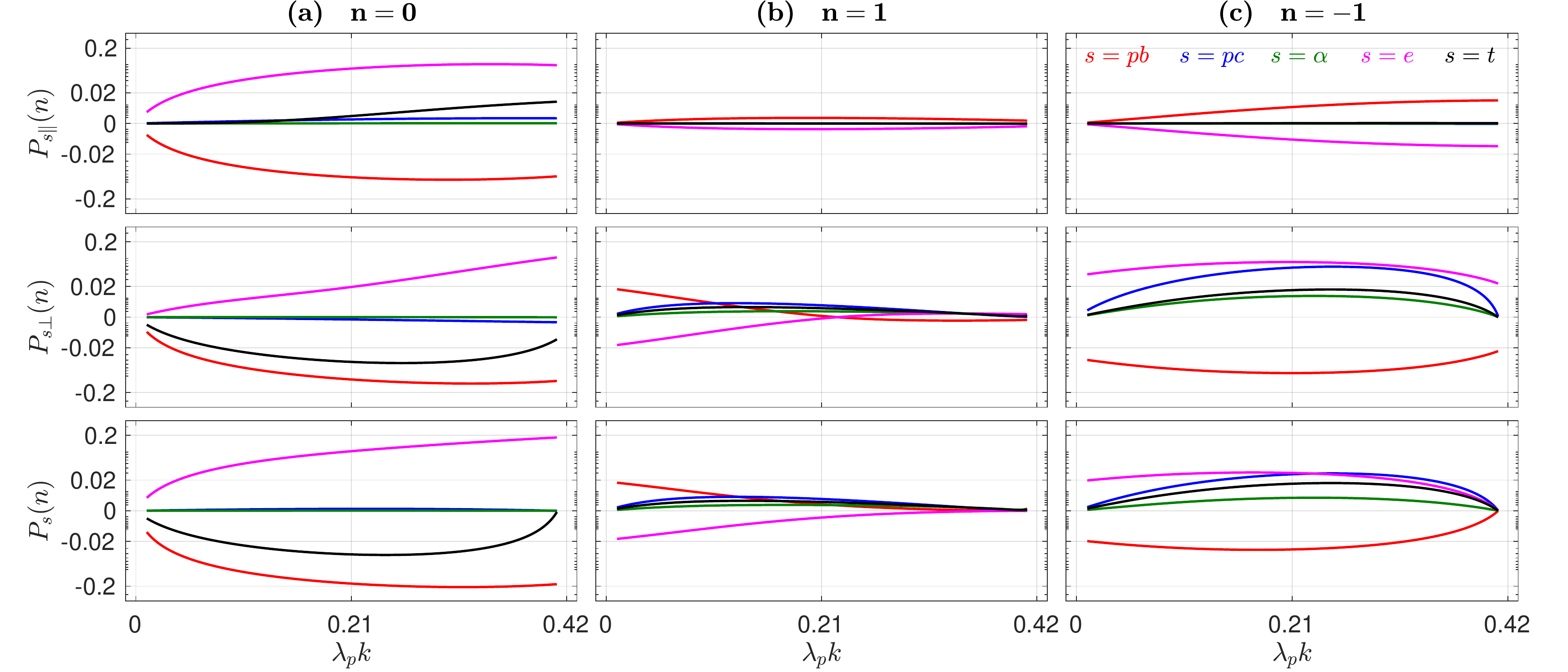}
\caption{
Energy transfer rates at different $n$ in the oblique fast-magnetosonic/whistler instability at point $\#2$: (a) $n=0$; (b) $n=1$; and (c) $n=-1$. (Upper panels) parallel energy transfer rates; (middle panels) perpendicular energy transfer rates; and (bottom panels) total energy transfer rates. 
\label{fig:1dot5va-33-N}}
\end{figure*}

\begin{figure*}
\includegraphics[width=\textwidth]{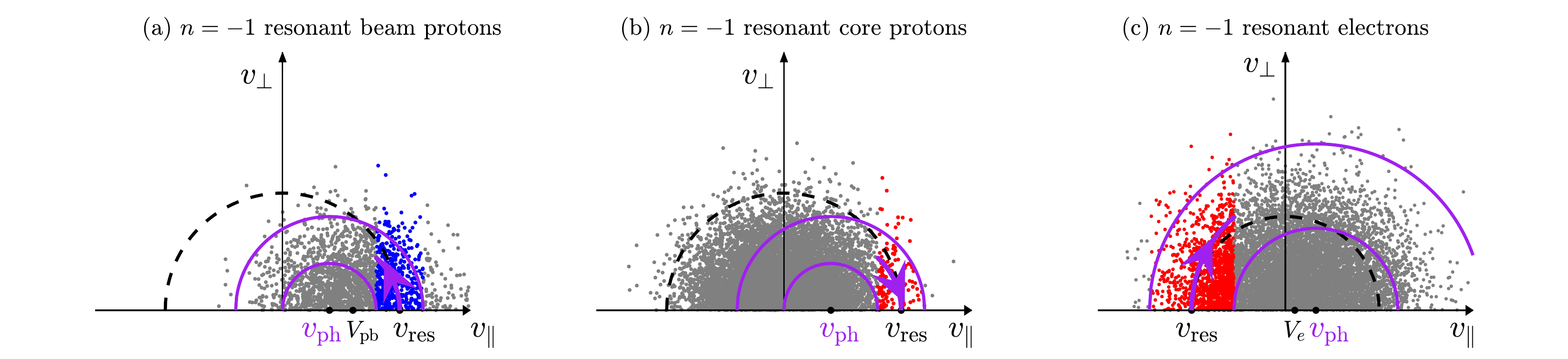}
\caption{
Diffusion paths in the oblique fast-magnetosonic/whistler instability. (a) Cyclotron interactions with $n=-1$ resonant beam protons with $v_{\mathrm{res}} - 2V_{\mathrm{Tp}} \leqslant v_{\parallel}\leqslant v_{\mathrm{res}} + 2V_{\mathrm{Tp}}$. (b) Cyclotron interactions with $n=-1$ resonant core protons with $v_{\mathrm{res}} - 2V_{\mathrm{Tp}} \leqslant v_{\parallel}\leqslant v_{\mathrm{res}} + 2V_{\mathrm{Tp}}$. (c) Cyclotron interactions with $n=-1$ resonant electrons with $v_{\mathrm{res}} - 2V_{\mathrm{Te}} \leqslant v_{\parallel}\leqslant v_{\mathrm{res}} + 2V_{\mathrm{Te}}$. Blue (red) points represent particles diffusing into a region with smaller (larger) $v$. The black dashed and purple solid curves denote $v_\parallel^2 + v_\perp^2 = C$ and $(v_{\parallel}-v_{\mathrm{ph}})^2+v_{\perp}^2=C$, respectively. The points are randomly sampled from the Maxwellian model for the proton beam, proton core, or electron population.
\label{fig:pathn=-1}}
\end{figure*}

Using the magnetic field and plasma parameters stated in Subsection~\ref{sub:2.2}, we find there are four typical proton beam instabilities in the inner heliosphere. An example is given in Figure~\ref{fig:k-theta_10Rs}, which presents the $k-\theta$ distributions of these typical instabilities driven by proton beams with $V_{\mathrm{pb}}=V_A$, $1.5V_A$, and $2.5V_A$ at $r=10R_S$. The $V_{\mathrm{pb}}=V_A$ proton beam drives an oblique Alfv\'en/ion-cyclotron instability, producing the left-hand polarized waves $\mathrm{arg}(E_y/E_x)=-90^\circ$, in which the maximum growth rate (marked by $\#1$) occurs at $\theta\simeq38^\circ$. The $V_{\mathrm{pb}}=1.5V_A$ beam results in an oblique fast-magnetosonic/whistler instability where the strongest excitation (marked by $\#2$) occurs at $\theta\simeq33^\circ$, and this instability generates the right-hand polarized waves with $\mathrm{arg}(E_y/E_x)=90^\circ$. The proton beam with a large speed $V_{\mathrm{pb}}=2.5V_A$ triggers two kinds of instabilities: an oblique Alfv\'en/ion-beam instability with the local maximum growth rates (point $\#3$ and $\#4$) at $\theta\simeq45^\circ$ and $\theta\simeq73^\circ$, and a parallel fast-magnetosonic/whistler instability with the maximum growth rate at $\theta=0^\circ$ ($\#5$). The nature of each instability is further explored in Figures~\ref{fig:1VA38degree}$-$\ref{fig:2dot5VA0degree}, which give a detailed analysis for each instability at its maximum growth rate.

\subsection{Oblique Alfv\'en/ion-cyclotron instability}

\begin{figure*}
\includegraphics[width=\textwidth]{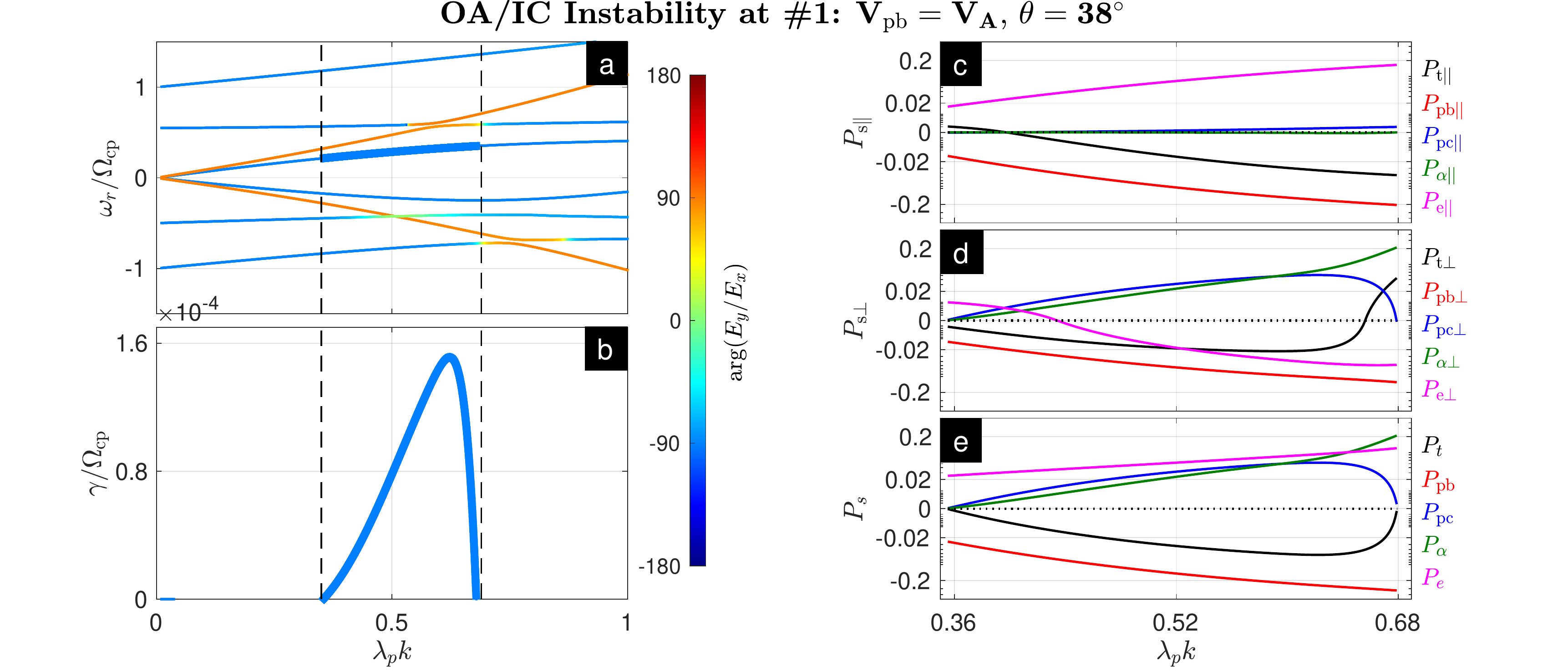}
\caption{
The OA/IC instability at point $\#1$ labeled in Figure 3: (a) the dispersion relations of all low-frequency waves; (b) the growth rate of the unstable wave; (c) parallel energy transfer rates; (d) perpendicular energy transfer rates; and (e) total energy transfer rates. The argument of $E_y/E_x$ is overlaid on the wave dispersion relation and growth rate. OA/IC = oblique Alfv\'en/ion-cyclotron.
\label{fig:1VA38degree}}
\end{figure*}

\begin{figure*}
\includegraphics[width=\textwidth]{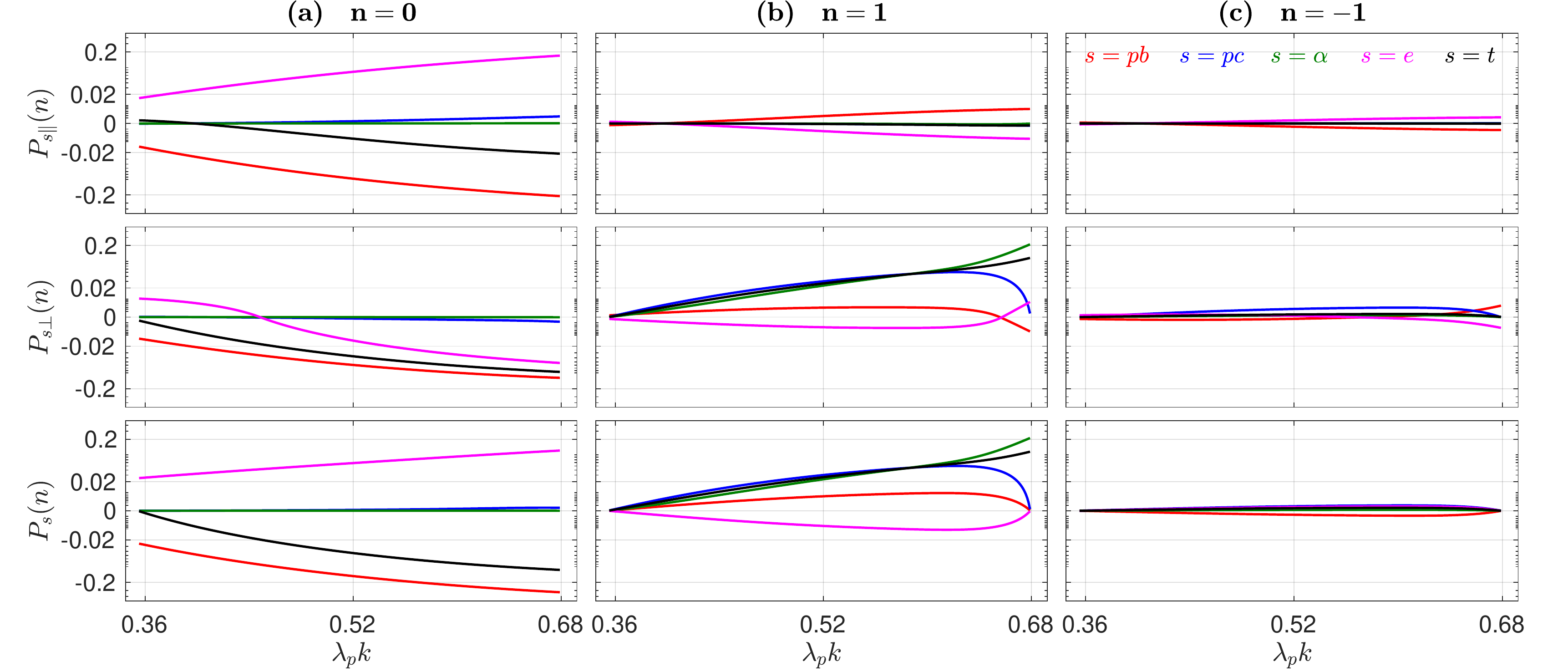}
\caption{
Energy transfer rates at different $n$ in the oblique Alfv\'en/ion-cyclotron instability: (a) $n=0$; (b) $n=1$; and (c) $n=-1$. (Upper panels) parallel energy transfer rates; (middle panels) perpendicular energy transfer rates; and (bottom panels) total energy transfer rates. 
\label{fig:1VA38-N}}
\end{figure*}

\begin{figure*}
\includegraphics[width=\textwidth]{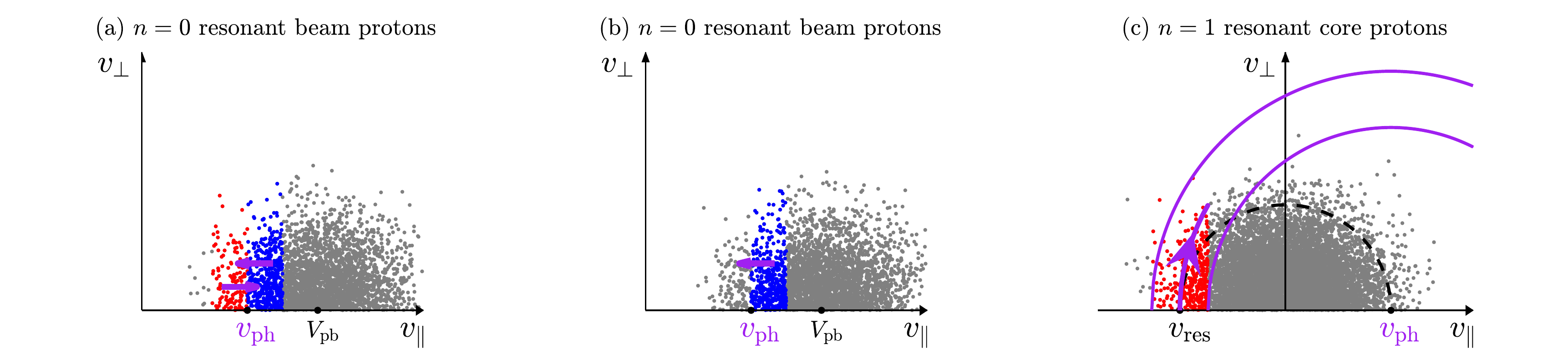}
\caption{
Diffusion paths in the oblique Alfv\'en/ion-cyclotron instability. (a) Landau interactions with $n=0$ resonant beam protons, where red and blue points denote protons with $v_{\mathrm{ph}} - 2V_{\mathrm{Tp}} \leqslant v_{\parallel}\leqslant v_{\mathrm{ph}}$ and $v_{\mathrm{ph}}  \leqslant v_{\parallel}\leqslant v_{\mathrm{ph}} + 2V_{\mathrm{Tp}}$, respectively. (b) Transit-time interactions with $n=0$ resonant beam protons with $v_{\mathrm{ph}} \leqslant v_{\parallel}\leqslant v_{\mathrm{ph}} + 2V_{\mathrm{Tp}}$. (c) Cyclotron interactions with $n=1$ resonant core protons with $v_{\mathrm{res}} - 2V_{\mathrm{Tp}} \leqslant v_{\parallel}\leqslant v_{\mathrm{res}} + 2V_{\mathrm{Tp}}$. Blue (red) points represent particles diffusing into a region with smaller (larger) $v$. The black dashed and purple solid curves denote $v_\parallel^2 + v_\perp^2 = C$ and $(v_{\parallel}-v_{\mathrm{ph}})^2+v_{\perp}^2=C$, respectively. The points are randomly sampled from the Maxwellian model for the proton beam or proton core population.
\label{fig:pathn=0and1}}
\end{figure*}

Figures~\ref{fig:1VA38degree}a and \ref{fig:1VA38degree}b present the dispersion relations of all low-frequency waves and the growth rate of the unstable wave under $V_{\mathrm{pb}}=V_A$ and $\theta=38^\circ$ at $r=10R_S$, respectively. Compared to the wave dispersion relations in Figure \ref{fig:WaveFluid}, it is evident that the unstable wave at point $\#1$ is the Alfv\'en/alpha-cyclotron mode wave. 

The energy transfer rates between unstable waves and particles are given in Figures~\ref{fig:1VA38degree}c$-$\ref{fig:1VA38degree}e. Through parallel electric field, the free energy of the proton beam is transferred into unstable waves. Simultaneously, the energy flows from unstable waves into the electron component; however, there is little energy transfer between unstable waves and the proton core (alpha particle) component in the parallel direction due to the parallel electric field. Since the total parallel energy transfer rate $P_{\mathrm{t\parallel}}$ is mainly smaller than zero, the net energy flows into unstable waves in the parallel direction. On the other hand, both proton beam and electron components can release energy to unstable waves in the perpendicular direction due to the perpendicular electric field, whereas both the proton core and alpha particle components gain energy from unstable waves. Since $P_{t\perp}\lesssim 0$, the unstable waves gain the net energy from particles in the perpendicular direction. From Figure \ref{fig:1VA38degree}e which presents the energy transfer rate sum of both the parallel and perpendicular directions, we see that the free energy carried by the proton beam is nearly equally flowing into unstable waves and other particle components at the position of the strongest instability. 

To understand the physical mechanisms driving parallel and perpendicular energy transfers, Figures~\ref{fig:1VA38-N} and \ref{fig:pathn=0and1} further show energy transfer rates at different $n$ and typical diffusive particle flux paths in the $n=0$ and $n=1$ resonances, respectively. Since the oblique Alfv\'en/ion-cyclotron instability generates oblique left-hand polarized waves, these waves mainly interact with particles through $n=0$ and $n=1$ resonance \citep[Figure~\ref{fig:1VA38-N}; also see][]{2013ApJ...764...88V}. 

A parallel electric field can induce strong Landau resonance interactions between unstable waves and $n=0$ resonant particles (Figure~\ref{fig:1VA38-N}a). 
Through comparing the wave dispersion relation and $n=0$ resonance lines $\omega_r= (V_{s}\pm 2V_{\mathrm{Ts}})k_\parallel$ and analyzing the diffusive particle flux path, we can estimate the effects of Landau resonance on unstable waves \citep[]{2013ApJ...764...88V}. 
The wave dispersion relation of the oblique Alfv\'en/ion-cyclotron wave, which can be approximately given as $\omega_{\mathrm{OA/IC}}\sim V_Ak_\parallel$, resides two resonance lines of the proton beam and electron components. This indicates that there are sufficient proton beam and electron particles taking part in Landau resonance interactions. 
For the proton beam population, $v_{\mathrm{ph}} \leqslant v_\parallel \leqslant v_{\mathrm{ph}} + 2V_{\mathrm{Tp}}$ and $ v_{\mathrm{ph}}-2V_{\mathrm{Tp}}\leqslant v_\parallel \leqslant v_{\mathrm{ph}}$ resonant protons may experience different diffusion paths (Figure~\ref{fig:pathn=0and1}a) that are similar to diffusion paths for the Landau resonant alpha particle beam population proposed by \cite{2013ApJ...764...88V}, and the total energy is flowing from these resonant protons into unstable waves, that is, $P_{\mathrm{pb\parallel}}(n=0)<0$. 
For the electron population, Landau interactions between resonant electrons with  $v_{\mathrm{ph}} - 2V_{\mathrm{Te}} \leqslant v_\parallel \leqslant v_{\mathrm{ph}} + 2V_{\mathrm{Te}}$ and unstable waves lead to $P_{\mathrm{e\parallel}}(n=0)>0$.
However, due to the wave dispersion relation highly deviating from two resonance lines of the proton core and alpha particle components, a few proton core and alpha particles experience Landau resonance interactions, and therefore $P_{\mathrm{pc\parallel}}(n=0)\sim 0$ and $P_{\mathrm{\alpha\parallel}}(n=0)\sim 0$. 

The $n=0$ resonant particles can suffer another kind of wave-particle interaction via a perpendicular electric field, that is, transit-time resonant interaction \citep[e.g.,][]{1992wapl.book.....S,1998ApJ...500..978Q}. Through the analysis for contributions of electric field components $E_x$ and $E_y$ on $P_{\mathrm{s\perp}}(n=0)$, we found that only  $E_y$ is responsible for  $P_{\mathrm{s\perp}}$ at $n=0$ (not shown). Because $E_y$ corresponds to $B_z$ via Faraday's law, $n=0$ resonant particles would be controlled by the motion equation $md_tv_\parallel=-\mu \widehat{\mathbf{b}}\cdot \nabla|\mathbf{B}|$, which will lead to resonant particles moving into smaller $v_\parallel$ \citep[Figure~\ref{fig:pathn=0and1}b; ][]{1992wapl.book.....S}, where $\mu=mv_\perp^2/2B_0$ is the magnetic moment of particles, and $\widehat{\mathbf{b}}=\mathbf{B}/B$. 
The transit-time resonant interaction leads $n=0$ resonant beam protons losing energy, i.e., $P_{\mathrm{pb\perp}}(n=0)<0$ (Figure~\ref{fig:1VA38-N}a).
Also, the transit-time resonant interaction can result in $n=0$ resonant electrons losing energy in the large $k$ region, i.e., $P_{\mathrm{e\perp}}(n=0)<0$ (Figure~\ref{fig:1VA38-N}a).

Furthermore, in order to satisfy $n=1$ cyclotron resonance condition $\omega_r=k_\parallel v_{\mathrm{res}}+\Omega_{\mathrm{cs}}$, resonant ions (electrons) should stream against (along) the background magnetic field. $n=1$ resonant ions can absorb energy from unstable waves, and $n=1$ resonant electrons can release energy into unstable waves (Figure~\ref{fig:1VA38-N}b). However, the contribution to $P_{\mathrm{t\perp}}(n=1)$ mainly comes from the proton core and alpha particle components. Figure~\ref{fig:pathn=0and1}c gives a sketch for the cyclotron resonance mechanism of core protons and alpha particles.
The quasi-linear diffusion theory predicts that resonant core protons and alpha particles are scattered along the surface of a constant kinetic energy in the wave frame \citep[]{1966PhFl....9.2377K,2013ApJ...764...88V}, i.e., $(v_{\mathrm{s\parallel}}-v_{\mathrm{ph}})^2+v_{\mathrm{s\perp}}^2=C$. When these resonant core protons and alpha particles move along the gradient in phase-space density (Figure~\ref{fig:pathn=0and1}c), they gain energy, as shown in Figure~\ref{fig:1VA38-N}b, inducing wave damping. 

It should be noted that a parallel electric field can lead to parallel energy transfer between unstable waves and $n=1$ resonant particles; however, this type interaction is weaker than the cyclotron resonant interaction.

\begin{figure*}
\includegraphics[width=\textwidth]{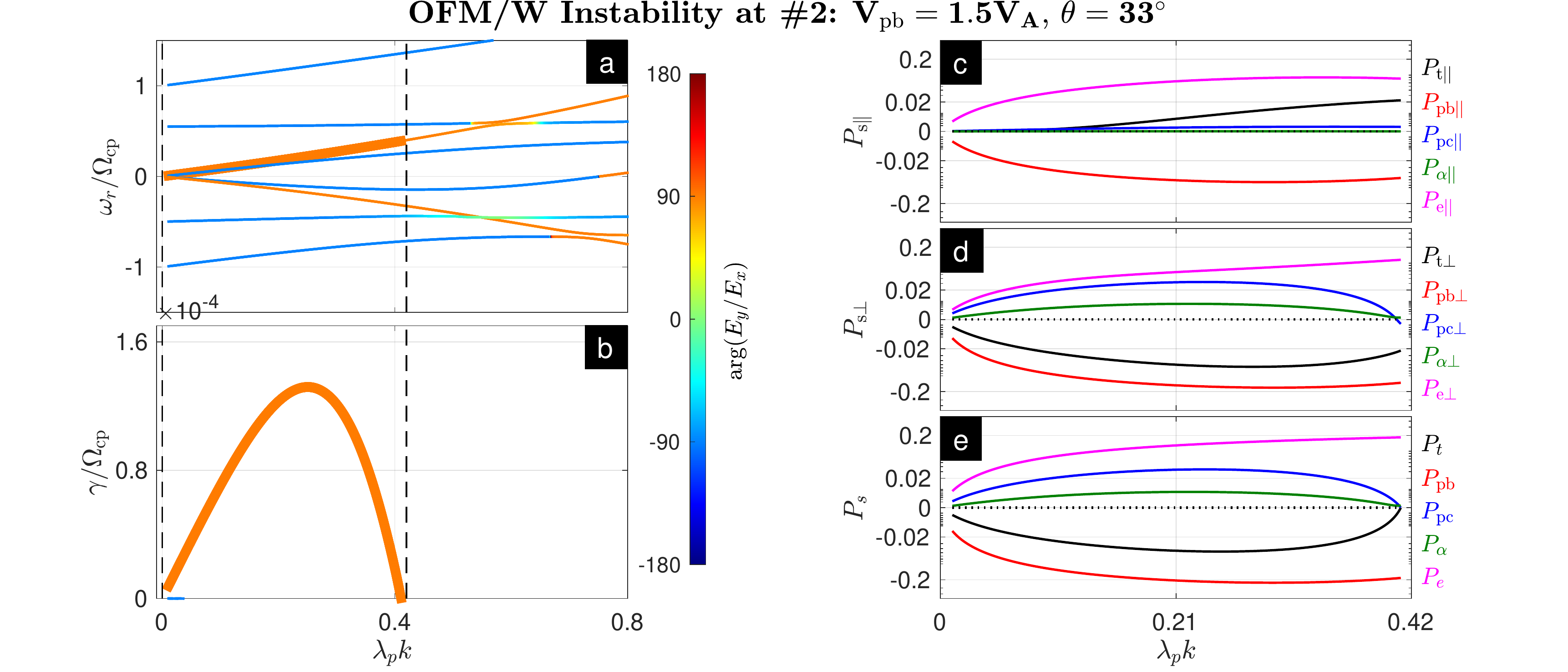}
\caption{
The OFM/W instability at point $\#2$. The description of panels (a)-(e) are the same as those in Figure~\ref{fig:1VA38degree}. OFM/W = oblique fast-magnetosonic/whistler.
\label{fig:1dot5VA33degree}}
\end{figure*}

\subsection{Oblique fast-magnetosonic/whistler instability}

\begin{figure*}
\includegraphics[width=\textwidth]{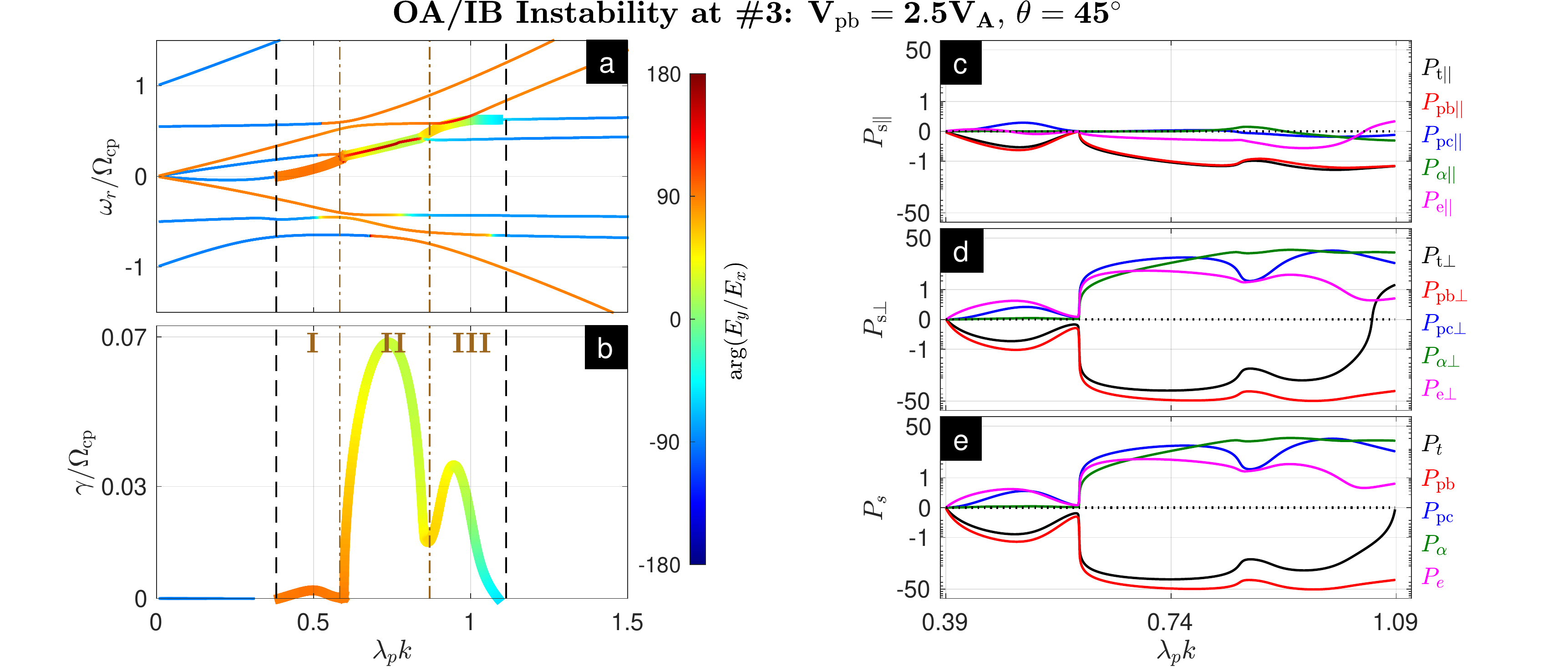}
\caption{
The OA/IB instability at point $\#3$. The instability can be classified into three types labeled by I, II, and III. The description of panels (a)-(e) are the same as those in Figure~\ref{fig:1VA38degree}. OA/IB = oblique Alfv\'en/ion-beam.
\label{fig:2dot5VA45degree}}
\end{figure*}

\begin{figure*}
\includegraphics[width=\textwidth]{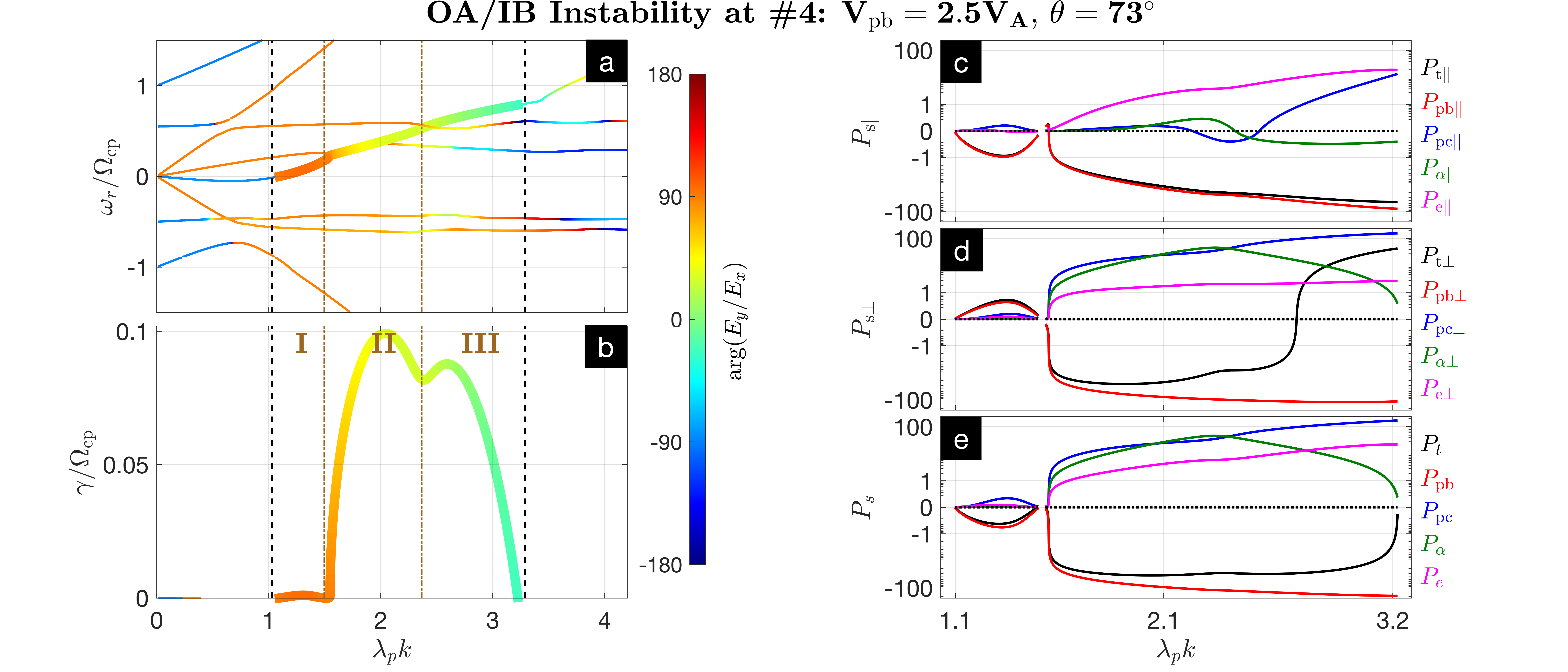}
\caption{
The OA/IB instability at point $\#4$. The instability can be classified into three types labeled by I, II, and III. The description of panels (a)-(e) are the same as those in Figure~\ref{fig:1VA38degree}. OA/IB = oblique Alfv\'en/ion-beam.
\label{fig:2dot5VA73degree}}
\end{figure*}

\begin{figure*}
\includegraphics[width=\textwidth]{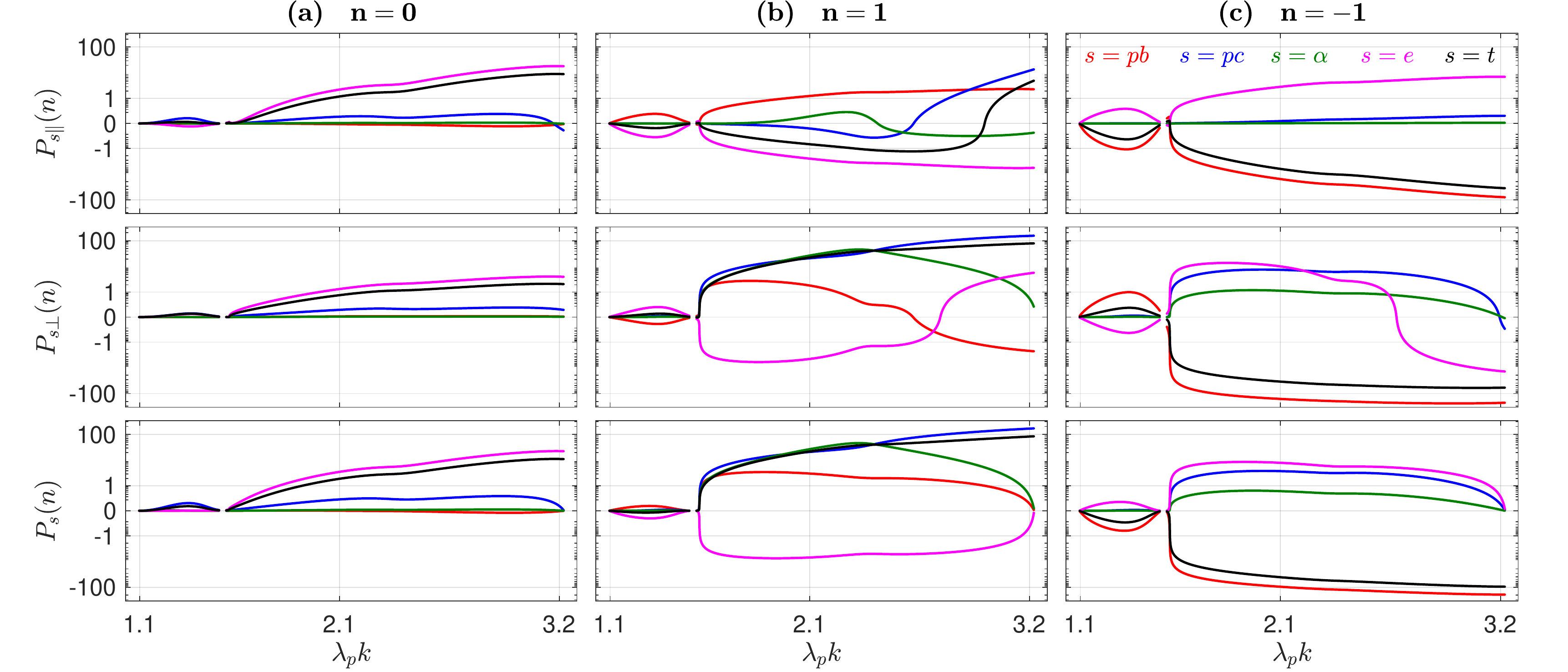}
\caption{
Energy transfer rates at different $n$ in the oblique Alfv\'en/ion-beam instability at point $\#4$: (a) $n=0$; (b) $n=1$; and (c) $n=-1$. (Upper panels) parallel energy transfer rates; (middle panels) perpendicular energy transfer rates; and (bottom panels) total energy transfer rates. 
\label{fig:2dot5va-73-N}}
\end{figure*}

Figures~\ref{fig:1dot5VA33degree}a and \ref{fig:1dot5VA33degree}b show all stable and unstable waves at point $\#2$, where $V_{\mathrm{pb}}=1.5V_A$ and $\theta=33^\circ$. We clearly see that the unstable wave at point $\#2$ corresponds to the fast-magnetosonic/whistler mode wave. 

Figures \ref{fig:1dot5VA33degree}c$-$\ref{fig:1dot5VA33degree}e present energy transfer rates between unstable waves and particles. These unstable waves absorb energy from the proton beam and mainly release energy toward the electron component in the parallel direction, in which the net energy flows from unstable waves into particles (Figure \ref{fig:1dot5VA33degree}c). In the perpendicular direction, shown in Figure \ref{fig:1dot5VA33degree}d, the net energy flowing into unstable waves approximates energy flowing into the electron component, and these two energy strengths are higher than that transferring into the proton core and alpha particle components. Consequently, the free energy lost from the proton beam flows into electrons, unstable waves, core protons, and alpha particles in sequence (Figure \ref{fig:1dot5VA33degree}e).  

Since oblique fast-magnetosonic/whistler waves have both parallel and perpendicular electric field fluctuations, these waves can interact with particles through $n=0$ and $n=-1$ resonances, as shown in Figure \ref{fig:1dot5va-33-N}, which gives energy transfer rates at different $n$. 

For $n=0$ resonant beam protons, their diffusive particle flux paths induced by a parallel electric field are the same as that illustrated in Figure \ref{fig:pathn=0and1}a, and therefore these protons lose the kinetic energy. $n=0$ resonant electrons gain energy from unstable waves through Landau resonance interactions. The perpendicular electric field induces energy flowing from $n=0$ resonant beam protons into unstable waves in the perpendicular direction, in which the mechanism is the same as that illustrate in Figure \ref{fig:pathn=0and1}b. However, different from $P_{\mathrm{e\perp}}\left(n=0\right)<0$ in Figure~\ref{fig:1VA38-N}a, $n=0$ resonant electrons obtain energy from unstable waves in the perpendicular direction, i.e., $P_{\mathrm{e\perp}}\left(n=0\right)>0$ in Figure~\ref{fig:1dot5va-33-N}a. This indicates that $n=0$ resonant electrons are scattered along the gradient in the electron phase-space density, that is, these electrons move to the region with higher kinetic energy. 

Under the resonance condition of $\omega_r=k_\parallel v_{\mathrm{res}}-\Omega_{\mathrm{cs}}$, $n=-1$ resonant particles correspond to forward streaming ions and backward streaming electrons. These particles should follow the diffusive particle flux paths illustrated in Figure~\ref{fig:pathn=-1}. As a consequence, beam protons lose the kinetic energy, and other particle components gain energy, resulting in $P_{\mathrm{pb}}(n=-1)< 0$ and $P_{\mathrm{pc,\alpha,e}}(n=-1)>0$. In addition, because oblique fast-magnetosonic/whistler waves are not pure right-hand polarized mode, i.e., $E_x+iE_y=0$ and $E_x-iE_y\neq0$, there also exists weak $n=1$ resonances, as shown in Figure \ref{fig:1dot5va-33-N}.

\subsection{Oblique Alfv\'en/ion-beam instability}

At points $\#3$ and $\#4$ in Figure~\ref{fig:k-theta_10Rs}c, the unstable wave corresponds to the oblique Alfv\'en/proton-beam mode wave, which is the reason why this instability is referred to as the oblique Alfv\'en/ion-beam instability. According to the dispersion relation and the growth rate of unstable waves in Figures \ref{fig:2dot5VA45degree} and \ref{fig:2dot5VA73degree}, this instability is further classified into three types: Type-I, Type-II, and Type-III. 

Here, we determine the basic features of these three type instabilities from Figures \ref{fig:2dot5VA45degree}a$-$\ref{fig:2dot5VA45degree}b, which exhibit unstable Alfv\'en/ion-beam mode waves at $V_{\mathrm{pb}}=2.5V_A$ and $\theta=45^\circ$. The unstable wave in Type-I instability corresponds to the low-frequency branch of the Alfv\'en/proton-beam wave, and this Type-I instability is the instability arising at $\#3$ in Figure \ref{fig:k-theta_10Rs}c. Different from Type-I instability where the Alfv\'en/proton-beam wave decouples with the Alfv\'en/alpha-cyclotron wave, these two mode waves are coupled in Type-II instability. Type-II instability was previously named as Alfv\'en I instability by \cite{1998JGR...10320613D}. Besides, Type-III instability arises in the region where the oblique Alfv\'en/proton-beam mode wave meets the alpha-cyclotron mode wave. 

Since Type-I is much weaker than Type-II and Type-III instabilities, energy transfer rates in Type-I instability are slower than that in other two instabilities. Here, we only discuss energy transfer rates in Type-II and Type-III instabilities. Figures \ref{fig:2dot5VA45degree}c$-$\ref{fig:2dot5VA45degree}e show energy transfer rates of unstable Alfv\'en/ion-beam mode waves at $V_{\mathrm{pb}}=2.5V_A$ and $\theta=45^\circ$. The energy transfer rate in the perpendicular direction dominates that in the parallel direction. In the perpendicular direction, beam protons lost energy in both Type-II and Type-III instabilities. However, there is an obvious difference between Type-II and Type-III instabilities, that is, core protons normally gain energy higher than alpha particles and electrons in Type-II instability, whereas alpha particles normally obtain energy approximating or slightly higher than core protons and electrons in Type-III instability. The reason for the latter is that Type-III instability excites alpha-cyclotron mode-like waves, which is in favor of the cyclotron resonance with alpha particles. 

For the oblique Alfv\'en/ion-beam instability at $\theta=73^\circ$ (see Figures \ref{fig:2dot5VA73degree}a$-$\ref{fig:2dot5VA73degree}e), the distributions of Type-I and Type-II instabilities are similar to that at $\theta=45^\circ$ in Figures \ref{fig:2dot5VA45degree}a$-$\ref{fig:2dot5VA45degree}e. Here, Type-II instability is the instability arising at point $\#4$ in Figure \ref{fig:k-theta_10Rs}. Type-III instability at $\theta=73^\circ$ is considerably different from that at $\theta=45^\circ$, and this instability produces the oblique Alfv\'en/proton-beam wave with $\omega\gtrsim \Omega_{\mathrm{c\alpha}}$. In Type-III instability, the energy lost from the proton beam is mainly flowing into core protons and alpha particles in the perpendicular direction, and unstable waves gain the net energy in both the parallel and perpendicular directions. 

Furthermore, Figure \ref{fig:2dot5va-73-N} shows energy transfer rates at different $n$ for unstable waves at $\theta=73^\circ$. It is interesting to see that energy transfer rates associated with $n=0$ resonant beam protons are nearly zero. Since the parallel phase velocity $\omega_{\mathrm{OA/IB}}/k_\parallel$ of unstable Alfv\'en/ion-beam mode waves approximates the drift speed of the proton beam, the total number of the proton beam particles with $v_{\parallel}>\omega_{\mathrm{OA/IC}}/k_\parallel$ is nearly the same as that with $v_{\parallel}<\omega_{\mathrm{OA/IB}}/k_\parallel$, and this could lead to no net energy transfer between the waves and beam protons.
Besides, because unstable waves are linearly polarized that means $E_x+iE_y\neq0$ and $E_x-iE_y\neq0$, both $n=1$ and $n=-1$ resonances are important, and the corresponding resonant mechanisms are similar to the $n=1$ resonance mechanisms shown in Figure \ref{fig:pathn=0and1} and the $n=-1$ resonance mechanisms shown in Figure \ref{fig:pathn=-1}.

\subsection{Parallel fast-magnetosonic/whistler instability}

\begin{figure}[t]
\centerline{
  \includegraphics[width=\columnwidth]{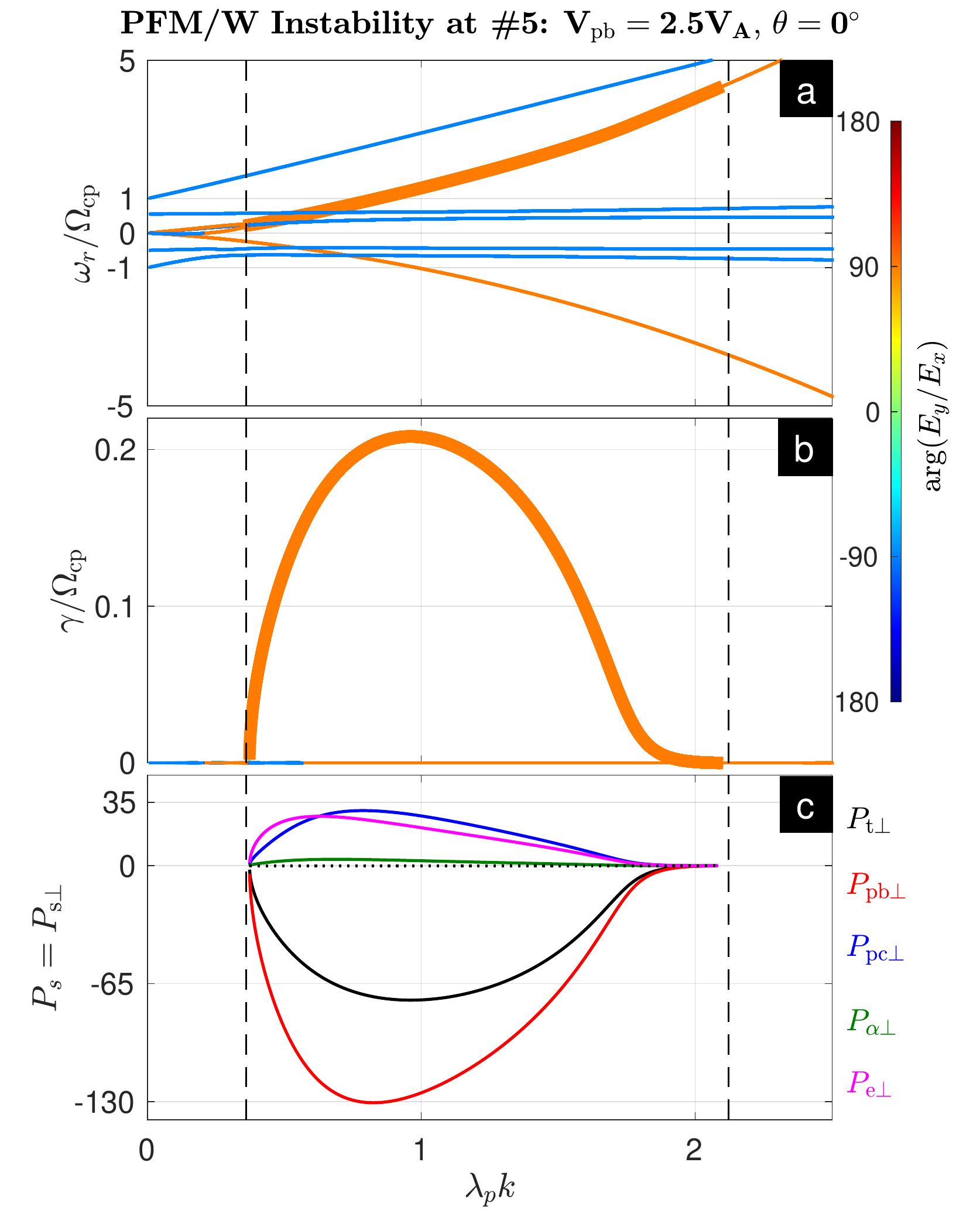}}
\caption{
The PFM/W instability at point $\#5$. The description of panels (a)-(c) are the same as those in Figure~\ref{fig:1VA38degree}.  PFM/W = parallel fast-magnetosonic/whistler.
\label{fig:2dot5VA0degree}}
\end{figure}

Parallel fast-magnetosonic/whistler instability generates parallel fast-magnetosonic/whistler waves at point $\#5$, as shown in Figures~\ref{fig:2dot5VA0degree}a and \ref{fig:2dot5VA0degree}b which present all low-frequency waves at $V_{\mathrm{pb}}=2.5V_A$ and $\theta=0^\circ$. Since the parallel fast-magnetosonic/whistler wave is a pure right-hand mode wave, only the $n=-1$ resonance exists, and the energy transfer is limited to the perpendicular direction, as shown in Figure \ref{fig:2dot5VA0degree}c. The diffusive particle flux paths of those resonant particles are the same as those given in Figure \ref{fig:pathn=-1}. As a result, beam protons lose energy, and other particle components gain energy.

\section{Radial distributions}

\begin{figure*}
\includegraphics[width=\textwidth]{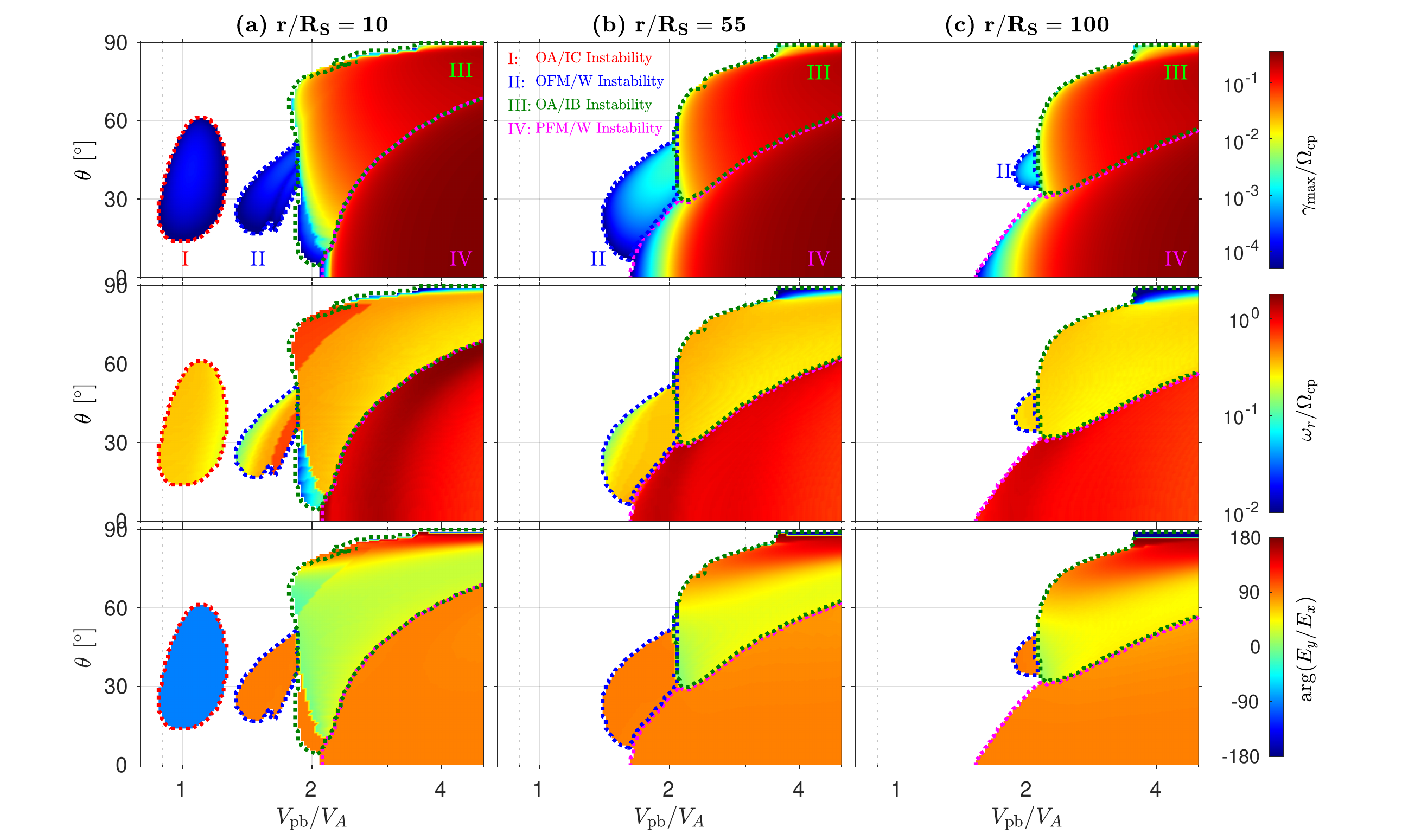}
\caption{
The $V_{\mathrm{pb}}-\theta$ distributions of proton beam instability at (a) $r=10R_S$, (b) $r=55R_S$, and (c) $r=100R_S$. (Top panels) The maximum growth rate, $\gamma_{\mathrm{max}}$; (second panels) the real frequency $\omega_r$ at $\gamma_{\mathrm{max}}$; and (bottom panels) the argument of $E_y/E_x$ at $\gamma_{\mathrm{max}}$. The regions controlled by OA/IC, OFM/W, OA/IB, and PFM/W instabilities are denoted by I, II, III, and IV, respectively. 
OA/IB = oblique Alfv\'en/ion-beam; OA/IC = oblique Alfv\'en/ion-cyclotron; OFM/W = oblique fast-magnetosonic/whistler; and PFM/W = parallel fast-magnetosonic/whistler.
\label{fig:theta-vpb}}
\end{figure*}

\begin{figure*}[t]
\includegraphics[width=\textwidth]{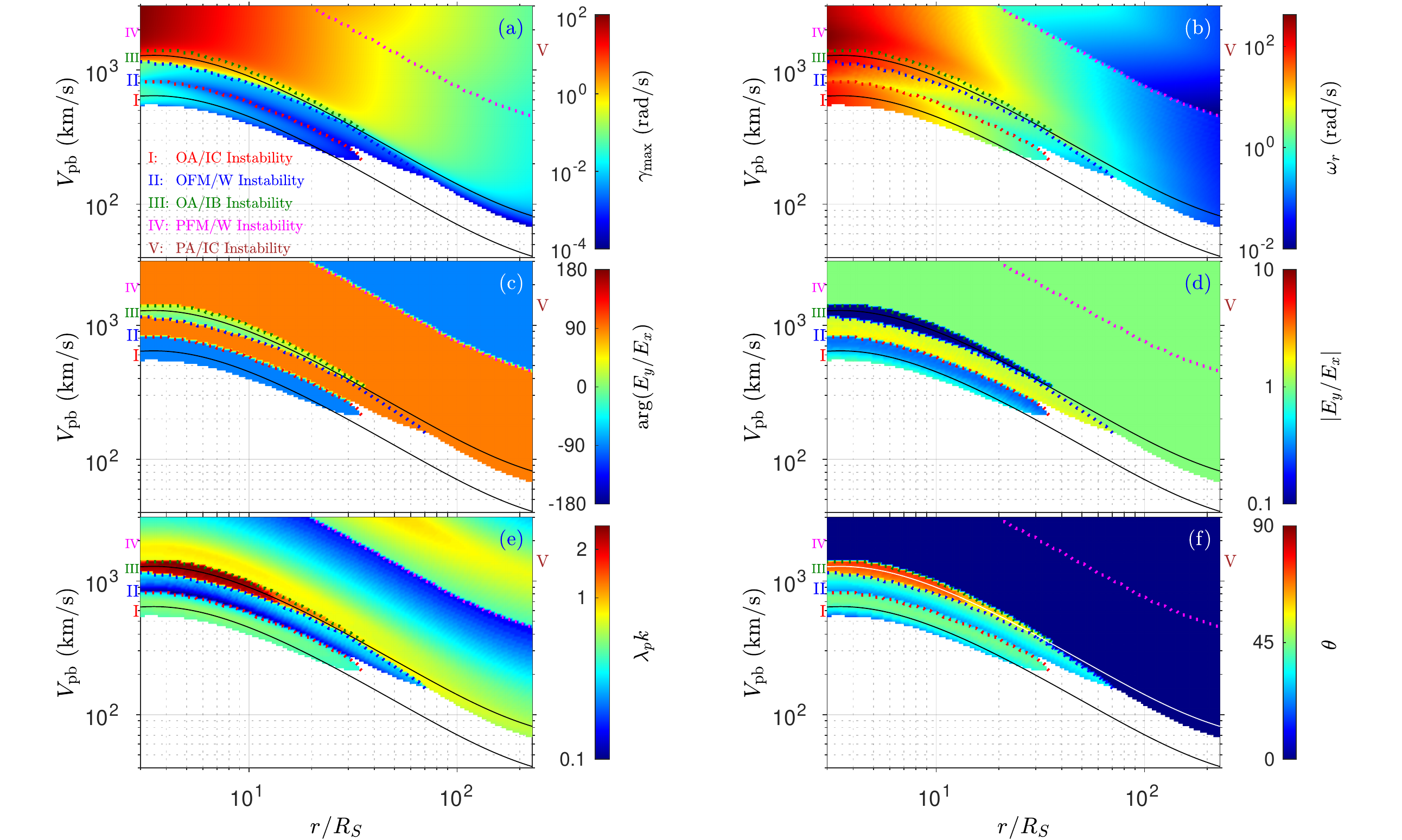}
\caption{
The $r-V_{\mathrm{pb}}$ distributions of proton beam instability: (a) the maximum growth rate, $\gamma_{\mathrm{max}}$; (b) the real frequency $\omega_r$ at $\gamma_{\mathrm{max}}$; (c) the argument of $E_y/E_x$ at $\gamma_{\mathrm{max}}$; (d) the magnitude of $E_y/E_x$ at $\gamma_{\mathrm{max}}$; (e) the wavenumber $k$ at $\gamma_{\mathrm{max}}$; and (f) the wave normal angle $\theta$ at $\gamma_{\mathrm{max}}$. The regions controlled by OA/IC, OFM/W, OA/IB, PFM/W and PA/IC instabilities are denoted by I, II, III, IV and V, respectively. 
The boundary between two kinds of instabilities is denoted by dotted curves. Two solid curves represent $V_{\mathrm{pb}}=V_A$ and $V_{\mathrm{pb}}=2V_A$, respectively.
OA/IB = oblique Alfv\'en/ion-beam; OA/IC = oblique Alfv\'en/ion-cyclotron; OFM/W = oblique fast-magnetosonic/whistler;  PA/IC=parallel Alfv\'en/ion-cyclotron; and PFM/W = parallel fast-magnetosonic/whistler.
\label{fig:r-vpb}}
\end{figure*}

\begin{figure*}
\includegraphics[width=\textwidth]{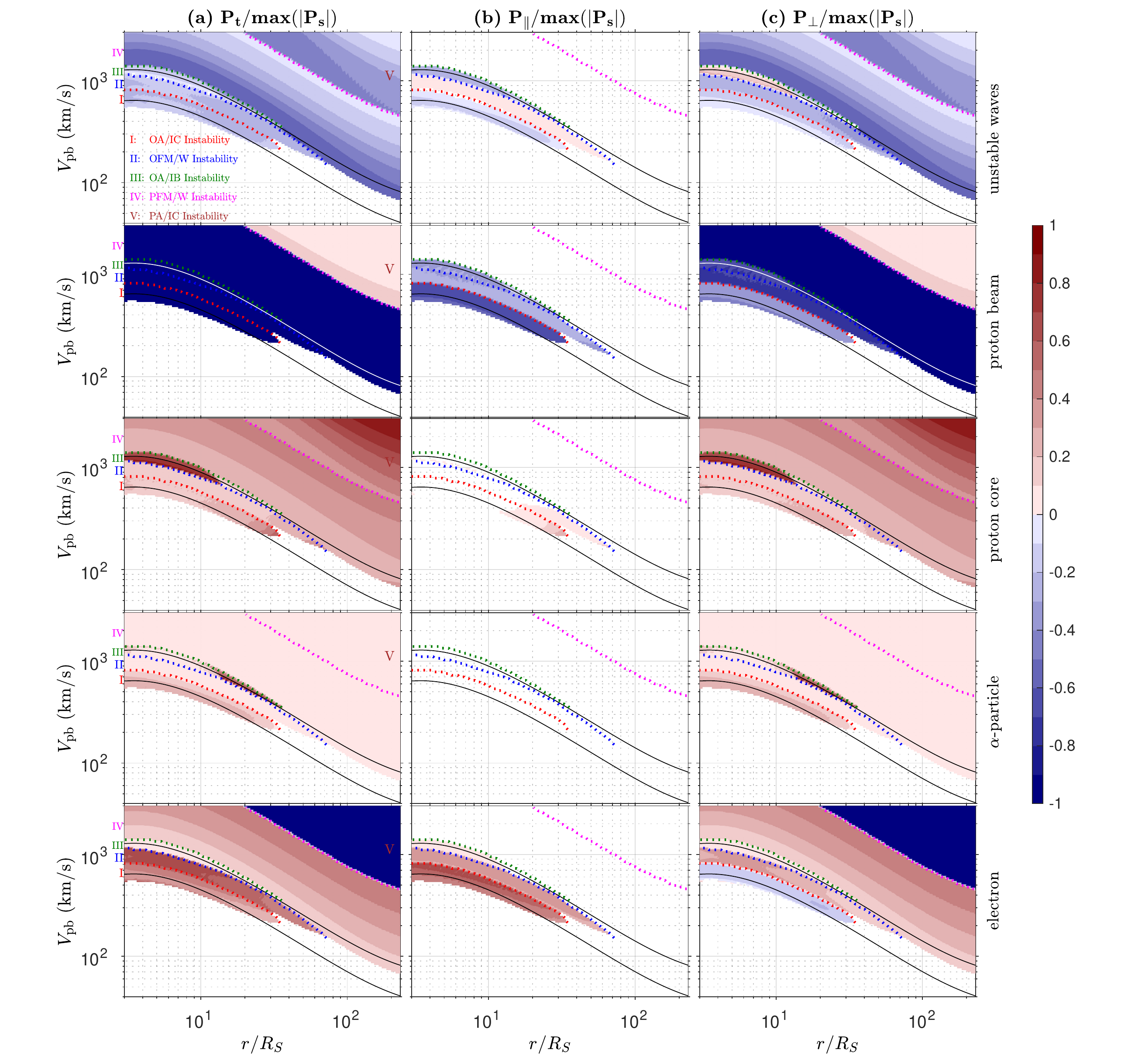}
\caption{
The $r-V_{\mathrm{pb}}$ distributions of (a) total, (b) parallel, and (c) perpendicular energy transfer rate in proton beam instability. These energy transfer rates are normalized by $\mathrm{max}(|P_s|)$ which corresponds to energy flowing from the instability source particles into unstable waves, e.g., $\mathrm{max}(|P_s|)=|P_{\mathrm{pb}}|$ in OA/IC, OFM/W, OA/IB, and PFM/W instabilities and $\mathrm{max}(|P_s|)=|P_{e}|$ in the PA/IC instability. The top panels give net energy transfer rates of unstable waves; and the second, third, fourth, and bottom panels present energy transfer rates associated with proton beam, proton core, alpha particle, and electron components, respectively. The regions controlled by OA/IC, OFM/W, OA/IB, PFM/W, and PA/IC instabilities are denoted by I, II, III, IV, and V, respectively. 
The boundary between two kinds of instabilities is denoted by dotted curves. Two solid curves represent $V_{\mathrm{pb}}=V_A$ and $V_{\mathrm{pb}}=2V_A$, respectively.
OA/IB = oblique Alfv\'en/ion-beam; OA/IC = oblique Alfv\'en/ion-cyclotron; OFM/W = oblique fast-magnetosonic/whistler;  PA/IC=parallel Alfv\'en/ion-cyclotron; and PFM/W = parallel fast-magnetosonic/whistler.
\label{fig:r-vpb-Ps}}
\end{figure*}

These four typical instabilities in Section 3 mainly control the evolution of the proton beam as it propagates outward from the Sun. In order to clearly explore controlling parameters of each instability, Figure \ref{fig:theta-vpb} presents the $V_{\mathrm{pb}}-\theta$ distributions of proton beam instability at three heliocentric distances: $r=10R_S$, $55R_S$, and $100R_S$. Oblique Alfv\'en/ion-cyclotron instability appears at $r=10R_S$ and disappears at $r=55R_S$ and $r=100R_S$. Oblique fast-magnetosonic/whistler instability is driven by proton beams with $V_{\mathrm{pb}}\sim1.3-1.8V_A$ at $r=10R_S$ and by larger $V_{\mathrm{pb}}$ beams at larger heliocentric distances. 
To understand these differences at different $r$, we checked the energy transfer rate of the Alfv\'en/alpha-cyclotron wave under $V_{\mathrm{pb}}=V_A$ and $\theta=38^\circ$ and of the fast-magnetosonic/whistler wave under $V_{\mathrm{pb}}=1.5V_A$ and $\theta=33^\circ$ at different heliocentric distances. For the oblique Alfv\'en/alpha-cyclotron wave, the Landau damping of core protons in the long-wavelength region and the cyclotron damping of alpha particles in the short-wavelength region significantly enhance at larger heliocentric distance. For the oblique fast-magnetosonic/whistler wave, Landau and cyclotron damping of core protons in the long-wavelength region and the cyclotron damping of electrons in the short-wavelength region considerably increase at larger heliocentric distance. Also, the energy transfer rate of the proton beam decreases with the heliocentric distance (this is due to the decrease of the proton beam speed). These effects result in different excitation behaviors of oblique Alfv\'en/ion-cyclotron and fast-magnetosonic/whistler instabilities at different heliocentric distances. 

Figure \ref{fig:theta-vpb} also shows that the normalized threshold $V_{\mathrm{pb}}/V_A$ of parallel fast-magnetosonic/whistler instability decreases with increasing heliocentric distance (with increasing $\beta_p$). However, the normalized threshold $V_{\mathrm{pb}}/V_A$ of oblique Alfv\'en/ion-beam instability increases with increasing heliocentric distance (with increasing $\beta_p$), and this instability can be always weaker than the parallel fast-magnetosonic/whistler instability at larger heliocentric distances.  
The $\beta_p$ dependence in these two instabilities is consistent with the results given by \cite{1976JGR....81.2743M} and \cite{1998JGR...10320613D}. 

The radial distribution of the ion-scale proton beam instability is shown in Figure \ref{fig:r-vpb}, which gives the maximum growth rate $\gamma_{\mathrm{max}}$, and the real frequency $\omega_r$, the argument angle of $E_y/E_x$, the absolute value of $E_y/E_x$, the wavenumber $\lambda_pk$ and the normal angle $\theta$ associated with $\gamma_{\mathrm{max}}$. Figure \ref{fig:r-vpb} exhibits different instabilities with different controlling regions. 
Oblique Alfv\'en/ion-cyclotron instability can control the region where $V_{\mathrm{pb}} \sim 0.8-1.4V_A$ and $r\lesssim 30R_S$.
Oblique fast-magnetosonic/whistler instability can arise in the region where $V_{\mathrm{pb}} \sim 1.3-2.0V_A$ and $r\lesssim 60R_S$.
Oblique Alfv\'en/ion-beam instability can exist in the region where $V_{\mathrm{pb}} \sim 1.7-2.2V_A$ and $r\lesssim 30R_S$. 
Parallel fast-magnetosonic/whistler instability mainly controls the instability of the proton beam having $V_{\mathrm{pb}} \sim 1.6-11V_A$. Besides, a parallel Alfv\'en/ion-cyclotron instability becomes dominant as the drifting speed of the proton beam is $V_{\mathrm{pb}} \gtrsim 11V_A$. One of important theoretical predictions is that unstable oblique Alfv\'en/ion-cyclotron and fast-magnetosonic/whistler waves can be produced by the proton beam in the solar atmosphere. 

Moreover, Figure \ref{fig:r-vpb-Ps} gives the $r-V_{\mathrm{pb}}$ distributions of the energy transfer rate. The basic features of the four typical instabilities are consistent with that explored in Section 3. For the fifth kind of instability, i.e., parallel Alfv\'en/ion-cycltron instability, backward drifting electrons lose the kinetic energy that provides the free energy to amplify unstable waves, and the three ion components gain energy from unstable waves.

\section{Effective excitation}

\begin{figure}[t]
\centerline{
  \includegraphics[width=8.5 cm]{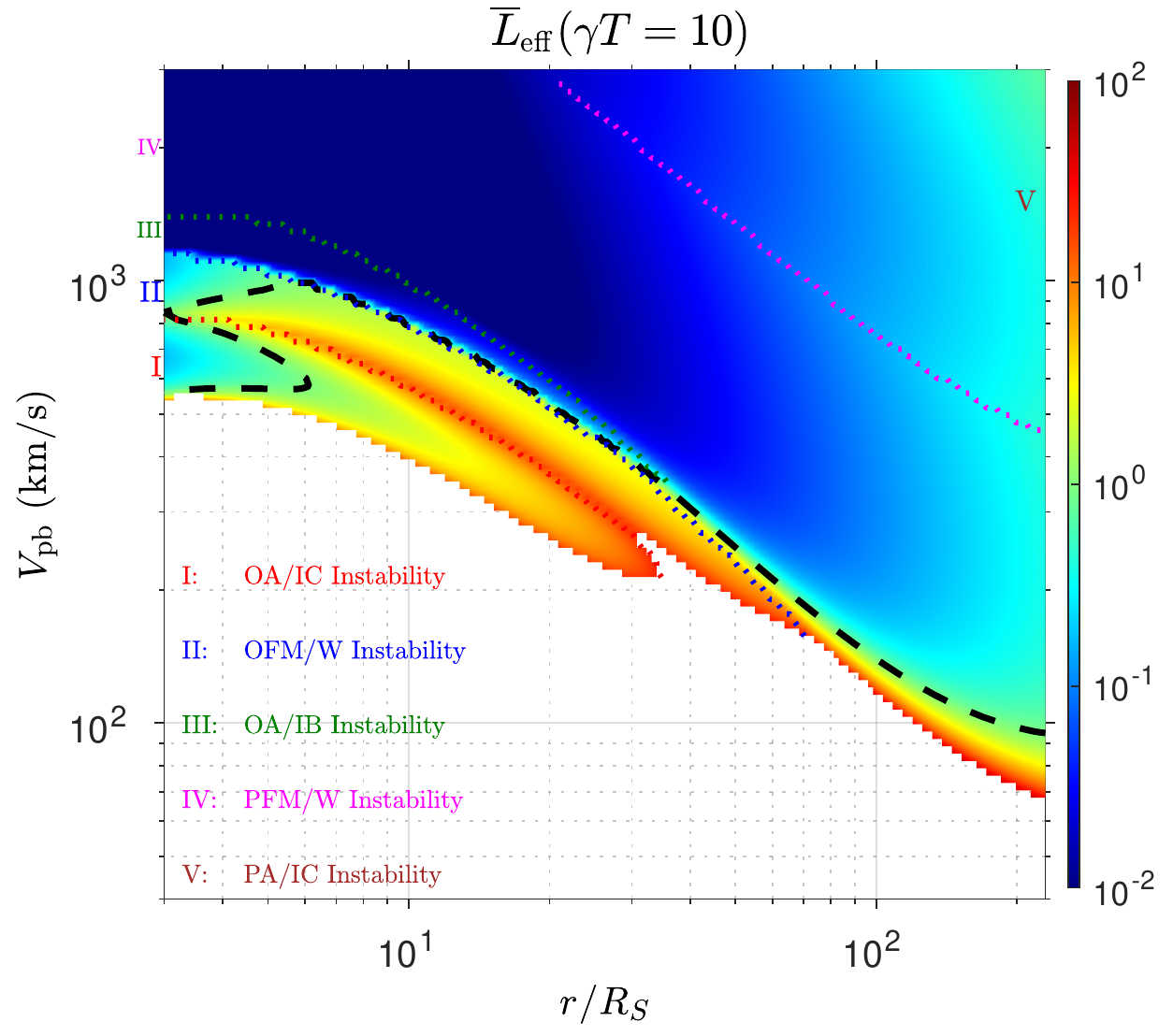}}
\caption{
The $r-V_{\mathrm{pb}}$ distribution of the normalized effective growth length $\bar{L}_{\mathrm{eff}}=L_{\mathrm{eff}}/R_S$. The dashed curve represents the counter line $\bar{L}_{\mathrm{eff}}=1$. The regions controlled by OA/IC, OFM/W, OA/IB, PFM/W, and PA/IC instabilities are denoted by I, II, III, IV, and V, respectively. 
The boundary between two kinds of instabilities is denoted by dotted curves. 
OA/IB = oblique Alfv\'en/ion-beam; OA/IC = oblique Alfv\'en/ion-cyclotron; OFM/W = oblique fast-magnetosonic/whistler; PA/IC = parallel Alfv\'en/ion-cycltron; and PFM/W = parallel fast-magnetosonic/whistler.
\label{fig:Leff}}
\end{figure}

Since the local Alfv\'en speed decreases with the heliocentric distance, when a proton beam is stable at small heliocentric distances, it will be destabilized at large heliocentric distances as its speed becomes larger than the local Alfv\'en speed therein.  \cite{2011JGRA..11611101H} have performed hybrid simulations to study the evolution of the proton beam in the expanding solar wind, and they indeed found a stable beam becomes unstable at large heliocentric distances. Since the energy carried by the proton beam is flowing into unstable waves and other particle components during proton beam instability, the proton beam is slowed down. Proton beam instability can lead to an effective constraint on the beam speed once unstable waves are considerably growing during a short time. 

To qualitatively estimate the effective excitation of the instability, we propose a parameter defined as the growth length $L_{\mathrm{grow}}$, which is the propagating distance of the proton beam during a period when unstable waves are linearly growing from a noise level to a considerable large amplitude. The growth length is expressed as 

\begin{equation}
L_{\mathrm{grow}} = \int_{t=t_{\mathrm{ini}}}^{t=t_{\mathrm{ini}}+T_{\mathrm{eff}}} \left( V_{\mathrm{sw}} + V_{\mathrm{pb}} \left(t\right) \right)dt,
\end{equation} 
where $T_{\mathrm{eff}}$ denotes the effective growing time. 
The corresponding wave amplitude $\delta B$ evolves as $\delta B=\delta B_{\mathrm{ini}}\mathrm{exp} \left( \int_{t=t_{\mathrm{ini}}}^{t=t_{\mathrm{ini}}+T_{\mathrm{eff}}} \gamma \left(t\right) dt \right)$ during $T_{\mathrm{eff}}$, where $\delta B_{\mathrm{ini}}$ is the wave amplitude at a noise level. Consequently, $L_{\mathrm{grow}}$ can be obtained once one knows $V_{\mathrm{pb}}\left(t\right)$ and $T_{\mathrm{eff}}$, in which $T_{\mathrm{eff}}$ are given as $\delta B/\delta B_{\mathrm{ini}}$ and $\gamma \left(t\right)$ are known. Since the free energy of the proton beam is continuously decreasing during the linear growth stage of the instability, both $V_{\mathrm{pb}}\left(t\right)$ and $\gamma \left(t\right)$ are decreasing. In principle, a quasi-linear theory should be used to proceed with a self-consistent treatment of the variables $V_{\mathrm{pb}}\left(t\right)$, $\gamma \left(t\right)$, and $\delta B/\delta B_{\mathrm{ini}}$. However, for performing a qualitative estimation, we consider a constant $\gamma$, and thus $T_{\mathrm{eff}}$ is given as $T_{\mathrm{eff}}=\mathrm{ln}(\delta B/\delta B_{\mathrm{ini}})/\gamma$. Assuming $\delta B/B_0=0.1$ and $\delta B_{\mathrm{ini}}/B_0=10^{-5}$, we have $T_{\mathrm{eff}}\simeq 9/\gamma$. 
On the other hand, we can also estimate $T_{\mathrm{eff}}$ through referring to the hybrid simulation results \citep[][]{1999JGR...104.4657D}, which exhibited that a transition from the linear growth stage to the nonlinear stage occurs at $\Omega_{\mathrm{cp}}t\sim 60$ in  parallel fast-magnetosonic/whistler instability and at $\Omega_{\mathrm{cp}}t\sim110$ in Alfv\'en I instability (oblique Alfv\'en/ion-beam instability). Considering the initial growth rate $\gamma=0.1\Omega_{\mathrm{cp}}$ \citep[][]{1999JGR...104.4657D}, $T_{\mathrm{eff}}$ is nearly $10/\gamma$. Furthermore, under the assumption of constant $V_{\mathrm{pb}}$, the growth length is estimated as 

\begin{equation}
L_{\mathrm{grow}} = \left( V_{\mathrm{sw}} + V_{\mathrm{pb}}\right) T_{\mathrm{eff}} = \left( V_{\mathrm{sw}} + V_{\mathrm{pb}}\right) \times 10/\gamma.
\end{equation} 

In fact, due to a highly variable plasma and magnetic field in the inner heliosphere, the plasma environment cannot always be favorable of a continuous development of proton beam instability. Here, we assume the characteristic spatial scale corresponds to one solar radius in the highly variable plasma environment. As a result, the effective excitation would be estimated by using the normalized growth length defined as ${\bar L}_{\mathrm{eff}} \equiv L_{\mathrm{grow}}/R_S$. ${\bar L}_{\mathrm{eff}}<1$ corresponds to an effective excitation, whereas ${\bar L}_{\mathrm{eff}}>1$ corresponds to an ineffective excitation. The radial distribution of ${\bar L}_{\mathrm{eff}}$ is presented in Figure \ref{fig:Leff}. It shows that oblique Alfv\'en/ion-cyclotron, oblique fast-magnetosonic/whistler and oblique Alfv\'en/ion-beam instabilities can be effectively excited by proton beams with $V_{\mathrm{pb}}\sim 600-1300$ km s$^{-1}$ in the $r\lesssim6R_S$ solar atmosphere, and parallel fast-magnetosonic/whistler instability can be effectively driven by proton beams with $V_{\mathrm{pb}}\gtrsim 2V_A$ in whole inner heliosphere.

\section{Discussion}

\subsection{Development of the plasma temperature anisotropy and its impact on  proton beam instability}
\label{sub:6.1}

\begin{figure*}[t]
\includegraphics[width=\textwidth]{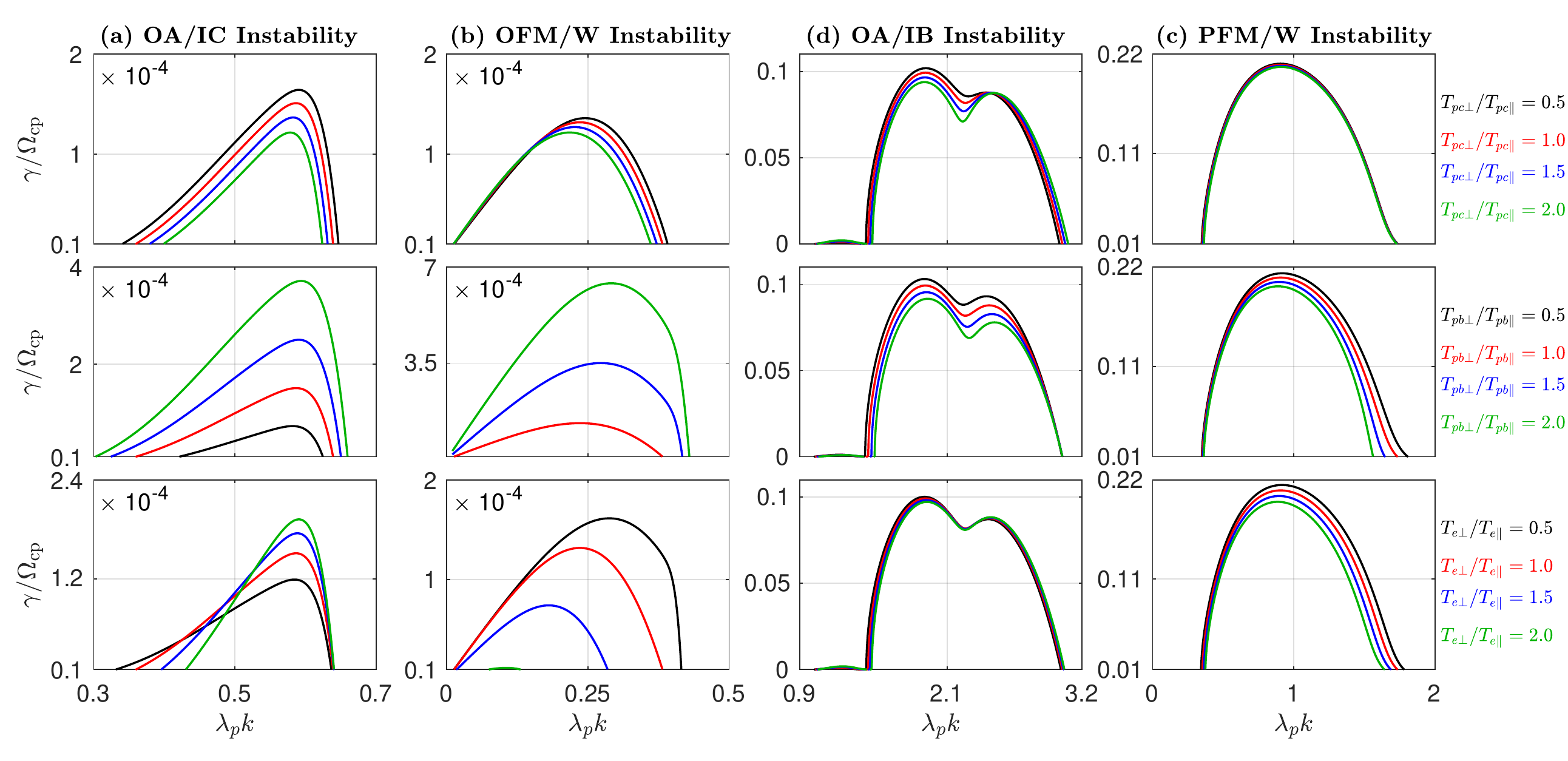}
\caption{
The dependence of the growth rate on the plasma temperature anisotropy in four typical proton beam instabilities at $r=10R_S$ shown in Figure 3: (a) the OA/IC instability; (b) the OFM/W instability; (c) the OA/IB instability; and (d) the PFM/W instability. The top, middle, and bottom panels give the dependence on the temperature anisotropy of the proton core, proton beam, and electron components, respectively. The black, red, blue, and green curves denote  $T_{\perp}/T_{\parallel}=0.5$, $1$, $1.5$, and $2$, respectively. 
OA/IB = oblique Alfv\'en/ion-beam; OA/IC = oblique Alfv\'en/ion-cyclotron; OFM/W = oblique fast-magnetosonic/whistler; and PFM/W = parallel fast-magnetosonic/whistler.
\label{fig:Tp_Depend}}
\end{figure*}

In this study, the plasma temperature is assumed to be isotropic for all particle species. Since energy transfer rates are different in the parallel and perpendicular directions, as shown in Figure~\ref{fig:r-vpb-Ps} and Table~\ref{table:parameters} in Appendix B, this will result in different developments of parallel and perpendicular temperatures once the instability is triggered.
Considering the energy transfer rates listed in Table~\ref{table:parameters} and assuming the energy totally responsible for the change in the plasma temperature, we can predict that:

(i) During oblique Alfv\'en/ion-cyclotron instability, due to $\overline{P}_{\mathrm{pc\perp}}\simeq 0.20\pm0.03$ and $\overline{P}_{\mathrm{pc\parallel}}\simeq 0.03\pm0.06$ for core protons, $T_{\mathrm{pc\perp}}$ will become larger than $T_{\mathrm{pc\parallel}}$. Similarly, $\overline{P}_{\mathrm{\alpha\perp}}\simeq0.19\pm0.06$ and $\overline{P}_{\mathrm{\alpha\parallel}}\simeq0$ will result in $T_{\mathrm{\alpha\perp}}/T_{\mathrm{\alpha\parallel}}>1$. From $\overline{P}_{\mathrm{e\perp}}\simeq - 0.14\pm0.06$ and $\overline{P}_{\mathrm{e\parallel}}\simeq0.52\pm0.07$, this implies that $T_{\mathrm{e\perp}}$ will decrease and $T_{\mathrm{e\parallel}}$ will increase, inducing $T_{\mathrm{e\perp}}/T_{\mathrm{e\parallel}}<1$. 

(ii) During oblique fast-magnetosonic/whistler instability, $\overline{P}_{\mathrm{pc\perp}}\simeq0.16\pm0.03$ and $\overline{P}_{\mathrm{pc\parallel}}\simeq0.02\pm0.02$ will result in $T_{\mathrm{pc\perp}}/T_{\mathrm{pc\parallel}}>1$. 
$\overline{P}_{\mathrm{\alpha\perp}}\simeq 0.04\pm0.01$ and $\overline{P}_{\mathrm{\alpha\parallel}}\simeq 0$ will induce $T_{\mathrm{\alpha\perp}}$ slightly larger than $T_{\mathrm{\alpha\parallel}}$. $\overline{P}_{\mathrm{e\perp}}\simeq 0.25\pm0.07$ and $\overline{P}_{\mathrm{e\parallel}}\simeq 0.32\pm0.02$ will cause $T_{\mathrm{e\perp}}/T_{\mathrm{e\parallel}}<1$. 

(iii) During oblique Alfv\'en/ion-beam instability, $T_{\mathrm{pc\perp}}/T_{\mathrm{pc\parallel}}>1$ and $T_{\mathrm{\alpha\perp}}/T_{\mathrm{\alpha\parallel}}>1$ will arise due to $\overline{P}_{\mathrm{pc\perp}}\simeq 0.56\pm0.26$, $\overline{P}_{\mathrm{pc\parallel}}\simeq0$, $\overline{P}_{\mathrm{\alpha\perp}}\simeq0.23\pm0.23$ and $\overline{P}_{\mathrm{\alpha\parallel}}\simeq0$. However, $T_{\mathrm{e\perp}}/T_{\mathrm{e\parallel}}$ will be unchanged because of $\overline{P}_{\mathrm{e\perp}}\simeq0.02\pm0.01$ and $\overline{P}_{\mathrm{e\parallel}}\simeq0.03\pm0.02$.

(iv) During parallel fast-magneosonic/whistler instability, $T_{\perp}/T_{\parallel}>1$ will arise in the proton core, alpha particle, and electron populations due to $\overline{P}_{\mathrm{pc\perp}}\simeq0.33\pm0.06$, $\overline{P}_{\mathrm{pc\parallel}}\simeq0$, $\overline{P}_{\mathrm{\alpha\perp}}\simeq0.04\pm0.01$, $\overline{P}_{\mathrm{\alpha\parallel}}\simeq0$, $\overline{P}_{\mathrm{e\perp}}\simeq0.30\pm0.10$ and $\overline{P}_{\mathrm{e\parallel}}\simeq0$.

(v) During parallel Alfv\'en/ion-cyclotron instability, $T_{\perp}/T_{\parallel}>1$ will arise in the proton beam, proton core, and alpha particle populations due to $\overline{P}_{\mathrm{pb\perp}}\simeq0.08\pm0.04$, $\overline{P}_{\mathrm{pb\parallel}}\simeq0$, $\overline{P}_{\mathrm{pc\perp}}\simeq0.54\pm0.16$, $\overline{P}_{\mathrm{pc\parallel}}\simeq0$, $\overline{P}_{\mathrm{\alpha\perp}}\simeq0.05\pm0.01$ and $\overline{P}_{\mathrm{\alpha\parallel}}\simeq0$.

The aforementioned predictions are performed for particle species that are not the source of each kind of instability. For the instability source particles, e.g., the proton beam in the first four instabilities and the electron beam population in the last instability, these particles will experience scattering by unstable waves along the velocity trajectory $\left(v_\parallel - v_{\mathrm{ph}} \right)^2 + v_\perp^2 = C$ in the wave frame, and the scattering of these particles can result in occurrence of the temperature anisotropy $T_\perp/T_\parallel>1$.
The development of $T_\perp/T_\parallel>1$ of the proton beam has been  identified in hybrid simulations of oblique Alfv\'en/ion-beam (Alfv\'en I) instability and parallel fast-magnetosonic/whistler instability \citep[e.g.,][]{1999JGR...104.4657D}. 

Actually, in situ satellites often detect anisotropic temperature in the solar wind plasma \citep[e.g.,][]{2002GeoRL..29.1839K,2006GeoRL..33.9101H,2018PhRvL.120t5102K,2020ApJS..246...70H}. The statistical analysis further showed that the temperature anisotropy changes with heliocentric distance \citep[e.g.,][]{2007GeoRL..3420105M}. The wave-particle interactions during and after proton beam instability in the solar wind can provide one source of the temperature anisotropy distribution therein \citep{1999JGR...104.4657D,2011JGRA..11611101H}. We note there still exist other sources, e.g., adiabatic expansion and Coulomb collisions. 

On the other hand, the temperature anisotropy will considerably affect proton beam instability \citep[e.g.,][]{1976JGR....81.2743M,1998JGR...10320613D,2019ApJ...884...44S,2020ApJ...899...61X}. 
Figure \ref{fig:Tp_Depend} exhibits the dependence of the four typical instabilities on $T_{\mathrm{pc\perp}}/T_{\mathrm{pc\parallel}}$, $T_{\mathrm{pb\perp}}/T_{\mathrm{pb\parallel}}$ and $T_{\mathrm{e\perp}}/T_{\mathrm{e\parallel}}$ in detail. This figure shows that the growth rate in oblique Alfv\'en/ion-cyclotron instability increases with increasing $T_{\mathrm{pb\perp}}/T_{\mathrm{pb\parallel}}$ and $T_{\mathrm{e\perp}}/T_{\mathrm{e\parallel}}$, and it decreases with increasing $T_{\mathrm{pc\perp}}/T_{\mathrm{pc\parallel}}$ before the temperature anisotropy instability is triggered \citep[see also][]{1976JGR....81.2743M}. For oblique Alfv\'en/ion-beam (Alfv\'en I) and parallel fast-magnetosonic/whistler instabilities, their growth rates decrease with increasing $T_\perp/T_\parallel$ of the proton beam, proton core, and/or electron species \citep[see also][]{1998JGR...10320613D}. 
Besides, Figure \ref{fig:Tp_Depend} shows that the growth rate in oblique fast-magnetosonic/whistler instability decreases with increasing $T_{\mathrm{pc\perp}}/T_{\mathrm{pc\parallel}}$ and $T_{\mathrm{e\perp}}/T_{\mathrm{e\parallel}}$, and this growth rate increases with increasing $T_{\mathrm{pb\perp}}/T_{\mathrm{pb\parallel}}$.

\subsection{Dependence of the alpha particle drift speed}

\begin{figure*}
\includegraphics[width=\textwidth]{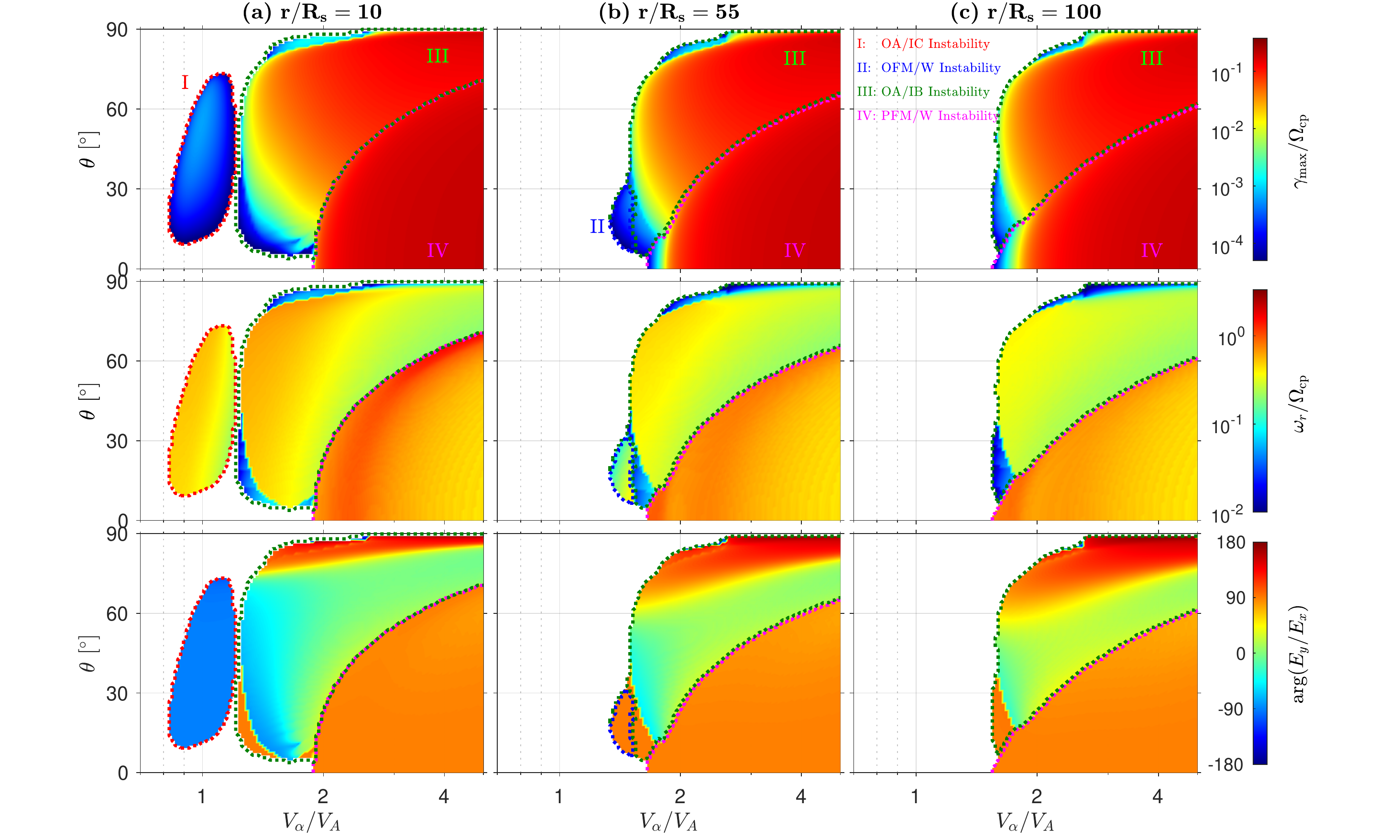}
\caption{
The $V_{\alpha}-\theta$ distributions of the alpha particle beam instability at (a) $r=10R_S$, (b) $r=55R_S$, and (c) $r=100R_S$. (Top panels) the maximum growth rate, $\gamma_{\mathrm{max}}$; (second panels) the real frequency $\omega_r$ at $\gamma_{\mathrm{max}}$; and (bottom panels) the argument of $E_y/E_x$ at $\gamma_{\mathrm{max}}$. The regions controlled by OA/IC, OFM/W, OA/IB, and PFM/W instabilities are denoted by I, II, III, and IV, respectively. 
OA/IB = oblique Alfv\'en/ion-beam; OA/IC = oblique Alfv\'en/ion-cyclotron; OFM/W = oblique fast-magnetosonic/whistler; and PFM/W = parallel fast-magnetosonic/whistler.
\label{fig:theta-vda}}
\end{figure*}

In addition to the proton beam, alpha particles also flow faster than core protons in the solar wind \citep[e.g.,][]{1982JGR....87...35M}. The differential drift between alpha and proton components can induce alpha particle beam instability \citep[e.g.,][]{2013ApJ...764...88V,2019ApJ...874..128L}. To show the difference between proton beam instability and alpha particle beam instability, Figure \ref{fig:theta-vda} presents the $V_{\alpha}-\theta$ distributions of the instability driven by alpha particle beams at three heliocentric distances, $r=10R_S$, $55R_S$ and $100R_S$. 
There are three typical instabilities at $r=10R_S$: Alfv\'en/ion-cyclotron instability, oblique Alfv\'en/ion-beam instability, and parallel fast-magnetosonic/whistler instability.
Also, oblique Alfv\'en/ion-beam instability arises at low $V_{\alpha}/V_A$ in comparison with the same kind of instability driven by the proton beam (see Figure \ref{fig:theta-vpb}). The detailed differences among different ion beam instabilities will be studied in the future.

\subsection{Dependence of the proton beam density}

\begin{figure*}
\includegraphics[width=\textwidth]{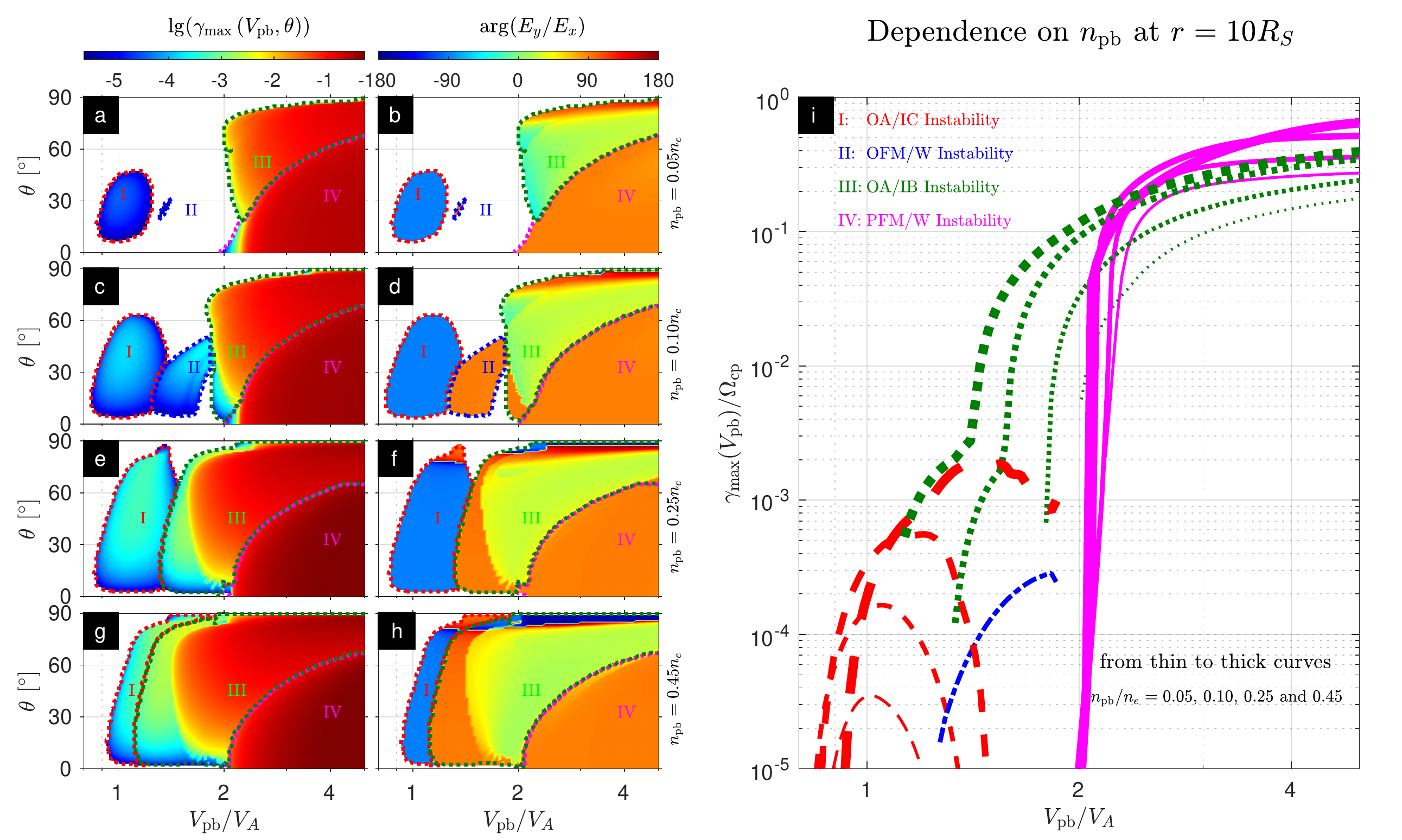}
\caption{
The dependence of proton beam instability on $n_{\mathrm{pb}}$ at $r=10R_S$: (a)$-$(b) $n_{\mathrm{pb}}=0.05n_e$; (c)$-$(d) $n_{\mathrm{pb}}=0.1n_e$; (e)$-$(f) $n_{\mathrm{pb}}=0.25n_e$;  and (g)$-$(h) $n_{\mathrm{pb}}=0.45n_e$. Panel (i) presents one-dimensional distribution of the maximum growth rate $\gamma_{\mathrm{max}}$ as a function of $V_{\mathrm{pb}}$.
The dashed, dashed-dotted, dotted, and solid curves represent OA/IC, OFM/W, OA/IB, and PFM/W instabilities, respectively. 
OA/IB = oblique Alfv\'en/ion-beam; OA/IC = oblique Alfv\'en/ion-cyclotron; OFM/W = oblique fast-magnetosonic/whistler; and PFM/W = parallel fast-magnetosonic/whistler.
\label{fig:np_r10}}
\end{figure*}

Proton beam instability is strongly dependent on the relative density of the proton beam \citep[e.g.,][]{1976JGR....81.2743M}. In order to show a comprehensive overview, the $V_{\mathrm{pb}}-\theta$ distributions of proton beam instability with four relative beam densities $n_{\mathrm{pb}}=0.05n_e$, $0.1n_e$, $0.25n_e$, and $0.45n_e$ are  presented in Figure \ref{fig:np_r10}, which exhibits the dependence of proton beam instability on $n_{\mathrm{pb}}/n_e$ at $r=10R_S$. When $n_{\mathrm{pb}}/n_e$ increases, the characteristic features in each instability are that: (1) the maximum growth rate in oblique Alfv\'en/ion-cyclotron instability arises at larger $\theta$; (2) oblique fast-magnetosonic/whistler instability is controlled by oblique Alfv\'en/ion-cyclotron instability or by oblique Alfv\'en/ion-beam instability; (3) oblique Alfv\'en/ion-beam instability becomes dominant at smaller $V_{\mathrm{pb}}/V_A$; and (4) the instability threshold of parallel fast-magnetosonic/whistler instability moves to larger $V_{\mathrm{pb}}/V_A$. 
These changes also occur at other heliocentric distances. It should be noted that although oblique Alfv\'en/ion-cyclotron instability is quenched at $r=55R_S$ and $100R_S$ in Figure \ref{fig:theta-vpb} where $n_{\mathrm{pb}}/n_e=0.1$, this instability can be driven by the proton beam with larger $n_{\mathrm{pb}}/n_e$, and the reason is that large $n_{\mathrm{pb}}/n_e$ provides more free energy for amplification of the instability, and small $n_{\mathrm{pc}}/n_e$ absorbs less energy from unstable waves (leading to weak damping).

Moreover, for a specific case, i.e., oblique Alfv\'en/ion-cyclotron instability at $V_{\mathrm{pb}}=V_A$ in low-beta plasma, the growth rate always increases with the relative beam density, which is consistent with results given in \cite{1976JGR....81.2743M}. For oblique Alfv\'en/ion-beam instability and parallel fast-magnetosonic/whistler instability at $V_{\mathrm{pb}}=2V_A$, their growth rates are first enhancing and then reducing as the relative proton beam density increases, and these two instabilities are even quenched at large relative beam densities \citep[]{1976JGR....81.2743M,1998JGR...10320613D}.

From Figure \ref{fig:np_r10}, we can see that proton beam instability are sensitive on $n_{\mathrm{pb}}/n_e$.  If we consider the value of $n_{\mathrm{pb}}/n_e$ different from $n_{\mathrm{pb}}/n_e=0.1$, the radial distribution of proton beam instability would be different from that shown in Figure \ref{fig:r-vpb}. However, the basic features are qualitatively consistent with that shown in Figures \ref{fig:r-vpb}. Recently, \cite{2020ApJS..248....5V} found that $n_{\mathrm{pb}}/n_{\mathrm{pc}}$ can be larger than 1 at $r\simeq 36R_s$, and this is more in favor of the excitation of proton beam instability in the inner heliosphere.

\subsection{Observational evidence of proton beam instability}

Since proton beam instability is widely thought of strongly constraining the proton beam in the solar-terrestrial environments, both statistical and case studies try to provide the observational evidence of the excitation of the instability \citep[e.g.,][]{1980JPlPh..23...91D,1986JGR....9113366L,1987JGR....92.7263M,2000GeoRL..27...53G,2016JGRA..121...30G,2019ApJ...883..185Z,2020ApJS..248....5V}. The statistical studies of  \cite{1987JGR....92.7263M} and \cite{2000GeoRL..27...53G} analyzed the $V_{\mathrm{pb}}-n_{\mathrm{pb}}$ distributions using Helios and Ulysses data, and they concluded that the beam instability constrains the proton beam speed in the solar wind. However, \cite{2004JGRA..109.5101T} found a weak correlation between Alfv\'en I instability and the observed data set in the $V_{\mathrm{pb}}-\beta_{\mathrm{pc}}$ distribution.  
Because \cite{2004JGRA..109.5101T} used the theoretical predictions of the Alfv\'en I instability under limited parameters \citep{1998JGR...10320613D}, the robust relation between the data set given by \cite{2004JGRA..109.5101T} and the theoretical results under the actual plasma parameters is still unclear. \cite{2018PhRvL.120t5102K} statistically studied the ion kinetic instability for 309 randomly selected events through Nyquist'’s instability criterion, and they found that a majority of the ion instabilities occur in the presence of a proton beam \citep[also see][]{2019ApJ...887..234K}. 

On the other hand, from case studies, the occurrence of proton beam instability is identified by previous works \citep[e.g.,][]{1980JPlPh..23...91D,1986JGR....9113366L,2016JGRA..121...30G,2019ApJ...883..185Z,2020ApJS..248....5V,2021ApJ...909....7K}. In particular, using PSP measurements, \cite{2020ApJS..248....5V} found coexistence of ion-scale waves and ion beams at $r\sim 40R_S$, and they identified the appearance of both parallel Alfv\'en/ion-cyclotron and fast-magnetosonic/whistler instabilities. Furthermore, \cite{2019ApJ...883..185Z} identified the interplay of proton beam and proton temperature anisotropy on proton instability, and they provided an observational evidence of proton instability enhanced by the proton beam. Recently, \cite{2021ApJ...909....7K} used two popular models (a single anisotropic population and two relatively drifting anisotropic populations) to fit the proton phase-space densities in two ion-scale wave activity events observed by PSP, and they found that the two-component model can result in instability much stronger than the one-component model.

The aforementioned works mainly checked the observations to the theoretical predictions of parallel fast-magnetosonic/whistler instability and oblique Alfv\'en/ion-beam (Alfv\'en I) instability. In this study, we propose that oblique Alfv\'en/ion-cyclotron instability and oblique fast-magnetosonic/whistler instability can be effectively excited in the solar atmosphere, which can be checked by using PSP measurements.

In addition to the resonant proton beam instabilities, a sufficiently dense and/or fast proton beam can provide sufficient excess parallel pressure to destabilize the CGL firehose instability \citep[e.g.,][]{2015JPlPh..81e3201K}. The instability condition is $\Lambda_f = \Sigma_s \left(\beta_{s\parallel}-\beta_{s\perp}\right)/2 + \Sigma_s \left(n_sm_sV_{s}^2\right)/\Sigma_s \left(n_sm_sV_A^2\right) >1$ \citep[e.g.,][]{2015JPlPh..81e3201K,2016ApJ...825L..26C}, which yields a relation of $V_{\mathrm{pb}} \gtrsim \sqrt{n_e/n_{\mathrm{pb}}}V_A\simeq 4.5V_A$ in the plasma with isotropic temperatures. However, in situ observations show that the proton beam with $V_{\mathrm{pb}} \gtrsim 4.5V_A$ may only exist upstream of the interplanetary shocks in the solar wind \citep[e.g.,][]{2017ApJ...849L..27K}.

\section{Summary}

This paper presents a comprehensive investigation of the energy transfer rate, radial distribution, and effective excitation of proton beam instability in the inner heliosphere. 

We firstly analyzed the nature and excitation mechanism of the four typical proton beam instabilities in Section 3, i.e., oblique Alfv\'en/ion-cyclotron instability, oblique fast-magnetosonic/whistler instability, oblique Alfv\'en/ion-beam instability, and parallel fast-magnetosonic/whistler instability. In particular, we find oblique Alfv\'en/ion-beam instability can be classified into three types (Figures \ref{fig:2dot5VA45degree} and \ref{fig:2dot5VA73degree}), and the wave mode in these three types corresponds to the long-wavelength branch of the Alfv\'en/ion-beam mode wave (Type-I), the coupling mode between the Alfv\'en/ion-beam mode and the Alfv\'en/ion-cyclotron mode (Type-II) and the coupling mode between the Alfv\'en/ion-beam mode and the alpha cyclotron mode (or the Alfv\'en/ion-beam mode wave at $\omega_r\gtrsim \Omega_{\mathrm{c\alpha}}$; Type-III). 

Based on the energy transfer rate and the diffusive particle flux path, we further clarified the roles of wave-particle resonant interactions on the instability excitation. Oblique Alfv\'en/ion-cyclotron instability is mainly triggered by Landau and transit-time interactions with $n=0$ resonant beam protons. Oblique fast-magnetosonic/whistler instability is produced through Landau and transit-time interactions with $n=0$ resonant beam protons and cyclotron interactions with $n=-1$ resonant beam protons. For oblique Alfv\'en/ion-beam instability, its excitation mechanisms include cyclotron interactions with $n=-1$ resonant beam protons and with $n=1$ resonant electrons as well as wave-particle interactions of these resonant particles resulting from the parallel electric field. In addition, parallel fast-magnetosonic/whistler instability is induced by the $n=-1$ cyclotron resonant interactions with beam protons.

Secondly, from the radial distribution of proton beam instability in the inner heliosphere, we exhibited the possible excitation region for each instability, i.e., the region with  $V_{\mathrm{pb}} \sim 0.8-1.4V_A$ and $r\lesssim 30R_S$ for oblique Alfv\'en/ion-cyclotron instability, the region with $V_{\mathrm{pb}} \sim 1.3-2V_A$ and $r\lesssim 60R_S$ for  oblique fast-magnetosonic/whistler instability, the region with $V_{\mathrm{pb}} \sim 1.7-2.2V_A$ and $r\lesssim 30R_S$ for oblique Alfv\'en/ion-beam instability, and parallel fast-magnetosonic/whistler instability controlling the whole inner heliosphere as $V_{\mathrm{pb}} \sim 1.6-11V_A$. It is evident that proton beam instability can provide a strong constraint on the proton beam speed in the inner heliosphere. We also exhibited the radial distributions of the energy transfer rate in these instabilities, which provide strong implications on different changes of plasma parallel and perpendicular temperatures in the inner heliosphere (see discussion in Subsection 6.1).

Furthermore, we proposed an effective excitation length to estimate the sufficient growth of proton beam instability in the inner heliosphere. In particular, we showed that oblique Alfv\'en/ion-cyclotron instability, oblique fast-magnetosonic/whistler instability and oblique Alfv\'en/ion-beam instability can be effectively excited by beam protons with the drift speed of $\sim 600-1300$ km s$^{-1}$. Since oblique Alfv\'en/ion-cyclotron and Alfv\'en/ion-beam waves can be significantly damped in the solar atmosphere, oblique Alfv\'en/ion-cyclotron instability and oblique Alfv\'en/ion-beam instability can contribute to the solar coronal heating during and after the instability. 

Lastly, this study shows the dependence of proton beam instability on the plasma parameters, such as the plasma temperature anisotropy, the drift speed of the alpha particles, and the relative density of the proton beam (see Figures \ref{fig:Tp_Depend}$-$\ref{fig:np_r10}). Although proton beam instability is indeed sensitive to these plasma parameters, our results explored the basic features of such instability.

\begin{deluxetable*}{ccccccccc}[th]
\tablenum{1}
\tablecaption{Typical parameters of proton beam instability$^\dagger$}
\tablewidth{0pt}
\tablehead{
\colhead{  } & \colhead{OA/IC} & \colhead{OFM/W} & \colhead{OA/IB} & \colhead{PFM/W}   & \colhead{PA/IC} 
}
%\decimalcolnumbers
\startdata
$r/R_S$ & $\lesssim 30$ & $\lesssim 60$   & $\lesssim 30$  & no limitation & no limitation     
\\
$V_{\mathrm{pb}}/V_A$ & $0.8-1.4$  
                                         & $1.3-2.0$  
                                         & $1.7-2.2$  
                                         & $1.6-11$  
                                         & $\gtrsim 11$    
\\
$\gamma/\Omega_{\mathrm{cp}}$ &$\sim 1.2\times10^{-4}$ %&$2\times10^{-5}\pm1\times 10^{-5}$
                                          & $\sim 3.4\times10^{-4}$%& $6\times10^{-5}\pm 5\times 10^{-5}$
                                          & $\sim3.2\times10^{-2}$%& $4.8\times10^{-3}\pm 2.8\times 10^{-3}$
                                          & $\sim0.3$% & $4.5\times10^{-2}\pm 1.6\times 10^{-2}$
                                          & $\sim0.7$%& $0.10\pm0.03$
\\
$\omega_r/\Omega_{\mathrm{cp}}$ & $\sim 0.30$  %& $0.05\pm 0.005$
                                            & $\sim 0.25$   %& $0.05\pm 0.005$  
                                            & $\sim 0.59$  %& $0.09\pm 0.02$ 
                                            & $\sim 0.84$    %& $0.13\pm 0.03$ 
                                            & $\sim0.52$   %& $0.08\pm 0.05$  
\\
$\mathrm{arg}(E_y/E_x)$ & $-90^\circ$ 
                                            & $90^\circ$
                                            & $\sim11^\circ$  
                                            & $90^\circ$
                                            & $-90^\circ$
\\
$|E_y/E_x|$                        & $0.42\pm0.15$ 
                                            & $1.55\pm0.14$ 
                                            & $0.07\pm0.03$ 
                                            & $1$ 
                                            & $1$ 
\\
$\lambda_pk$                    & $0.47\pm0.06$  
                                            & $0.23\pm0.07$   
                                            & $2.15\pm0.62$ 
                                            & $0.46\pm0.21$  
                                            & $0.63\pm0.13$   
\\
$\theta$                             & $38^\circ\pm7^\circ$   
                                            & $34^\circ\pm6^\circ$ 
                                            & $68^\circ\pm6^\circ$ 
                                            & $0^\circ$     
                                            & $0^\circ$     
\\
$\overline{P}_{\mathrm{pb}}$ & $-1$ & $-1$   & $-1$   & $-1$  & $0.08\pm0.04$ 
\\
$\overline{P}_{\mathrm{pb\perp}}$ & $-0.36\pm0.04$ 
                         & $-0.71\pm0.01$ 
                         & $-0.72\pm0.07$ 
                         & $-1$ 
                         & $0.08\pm0.04$ 
\\
$\overline{P}_{\mathrm{pb\parallel}}$ & $-0.64\pm0.04$ 
                              & $-0.29\pm0.01$  
                              & $-0.28\pm0.07$ 
                              & $0$   
                              & $0$  
\\
$\overline{P}_{\mathrm{pc}}$ & $0.23\pm0.08$ 
                & $0.17\pm0.05$ 
                & $0.56\pm0.27$ 
                & $0.33\pm0.07$ 
                & $0.54\pm0.16$  
\\
$\overline{P}_{\mathrm{pc\perp}}$ & $0.20\pm0.03$ 
                         & $0.16\pm0.03$  
                         & $0.56\pm0.26$   
                         & $0.33\pm0.06$
                         & $0.54\pm0.16$  
\\
$\overline{P}_{\mathrm{pc\parallel}}$ & $0.03\pm0.06$ 
                              & $0.02\pm0.02$  
                              & $0$    
                              & $0$   
                              & $0$   
\\
$\overline{P}_{\alpha}$ & $0.19\pm0.06$ 
                       & $0.04\pm0.01$ 
                       & $0.23\pm0.23$ 
                       & $0.04\pm0.01$ 
                       & $0.05\pm0.01$ 
\\
$\overline{P}_{\mathrm{\alpha\perp}}$ & $0.19\pm0.06$ 
                                & $0.04\pm0.01$ 
                                & $0.23\pm0.23$   
                                & $0.04\pm0.01$ 
                                & $0.05\pm0.01$ 
\\
$\overline{P}_{\mathrm{\alpha\parallel}}$ & $<0.001$ 
                                    & $<0.001$ 
                                    & $\sim0.001$ 
                                    & $0$ 
                                     & $0$ 
\\
$\overline{P}_{e}$ & $0.38\pm0.08$ 
              & $0.57\pm0.08$  
              & $0.05\pm0.02$  
              & $0.30\pm0.10$  
              & $-1$  
\\
$\overline{P}_{\mathrm{e\perp}}$ & $-0.14\pm0.06$ 
                       & $0.25\pm0.07$ 
                       & $0.02\pm0.01$  
                       & $0.30\pm0.10$  
                       & $-1$ 
\\
$\overline{P}_{\mathrm{e\parallel}}$ & $0.52\pm0.07$
                            & $0.32\pm0.02$ 
                            & $0.03\pm0.02$ 
                            & $0$ 
                            & $0$ 
\\
$\overline{P}_{t}$ & $-0.20\pm0.03$ 
              & $-0.22\pm0.04$  
              & $-0.16\pm0.09$  
              & $-0.34\pm0.17$  
              & $-0.33\pm0.12$  
\\
$\overline{P}_{\mathrm{t\perp}}$ & $-0.11\pm0.04$ 
                       & $-0.26\pm0.04$ 
                       & $0.08\pm0.13$  
                       & $-0.34\pm0.17$  
                       & $-0.33\pm0.12$ 
\\
$\overline{P}_{\mathrm{t\parallel}}$ & $-0.09\pm0.04$
                                            & $0.04\pm0.01$ 
                                            & $-0.24\pm0.05$ 
                                            & $0$ 
                                            & $0$ 
\\
\enddata 
\tablecomments{$\dagger$ $\overline{P}$ is defined as $\overline{P} \equiv P/\mathrm{max}(|P_s|)$, where $\mathrm{max}(|P_s|)$ is the magnitude of the energy transfer rate between unstable waves and instability source particles that provide free energy to trigger the instability. $|E_y/E_x|$, $\lambda_pk$, $\theta$ and $\overline{P}$ are given by using the mean and the standard deviation.}
\label{table:parameters}
\end{deluxetable*}

\begin{acknowledgments}
This work was supported by the NNSFC grant Nos. 41974203, 41531071, and 11673069. The author J.Z. appreciates the referee for helpful suggestions and inspiring comments.
\end{acknowledgments}

\appendix

\section{Several expressions of the energy transfer rate}

From Ampere's and Faraday's laws in Fourier space,
\begin{eqnarray}
\mathbf{k}\times \mathbf{B} &=& -i \mu_0 \mathbf{J} - \frac{\omega \mathbf{E}}{c^2},\\
\mathbf{k}\times \mathbf{E} &=& \omega \mathbf{B},
\end{eqnarray}
the total energy transfer rate can be given as
\begin{equation}
P_t = \frac{ \mathbf{E}\cdot \mathbf{J}^* + \mathbf{E}^*\cdot \mathbf{J}}{4W_{\mathrm{EB}}}=-2\gamma,
\label{Pt}
\end{equation}
which clearly indicates that $P_t$ and $\gamma$ have reversed energy flow directions.

In order to illustrate the difference between our expression of the energy transfer rate and the expression proposed in previous studies \citep[e.g.,][]{1998ApJ...500..978Q}, we rewrite $P_s$ in terms of the susceptibility tensor ${\bm \chi_s}$ using ${\bm \sigma_s}=-i\epsilon_0\omega{\bm \chi_s}$ and $\omega=\omega_r+i\gamma$,
\begin{eqnarray}
P_s = \omega_r \frac{\epsilon_0 \mathbf{E}^*\cdot {\bm \chi_{sa}} \cdot \mathbf{E}} {2W_{\mathrm{EB}}} 
+ \gamma \frac{\epsilon_0 \mathbf{E}^*\cdot{\bm \chi_{sh}} \cdot \mathbf{E}} {2W_{\mathrm{EB}}} ,
\label{Ps_A1}
\end{eqnarray}
where ${\bm \chi_{sh}}=\left( {\bm \chi_s} + {\bm \chi_s}^{\dag} \right)/2$ and ${\bm \chi_{sa}}=\left( {\bm \chi_s} - {\bm \chi_s}^{\dag} \right)/2i$ denote the Hermitian and anti-Hermitian parts of ${\bm \chi_s}$, respectively. Under the assumption of $\omega_r \gg \gamma$, Equation (\ref{Ps_A1}) can be reduced to the following expression at the condition of $\mathbf{E}^*\cdot {\bm \chi_{sa}} \cdot \mathbf{E} \gtrsim \mathbf{E}^*\cdot {\bm \chi_{sh}} \cdot \mathbf{E}$, 
\begin{eqnarray}
P_s \simeq \omega_r \frac{\epsilon_0 \mathbf{E}^*\cdot {\bm \chi_{sa}} \cdot \mathbf{E}} {2W_{\mathrm{EB}}}.
\label{Ps_A2}
\end{eqnarray}

Different from expressions (\ref{Ps_A1}) and (\ref{Ps_A2}), previous studies often use the energy transfer defined by \cite{1998ApJ...500..978Q} ,
\begin{eqnarray}
P^{Q}_s \equiv  \frac{\mathbf{E}^*\cdot {\bm \chi_{sa}} \cdot \mathbf{E}} {4W_{\mathrm{EB}}} ~[\mathrm{CGS}] = \frac{\pi \epsilon_0 \mathbf{E}^*\cdot {\bm \chi_{sa}} \cdot \mathbf{E}} {W_{\mathrm{EB}}}~[\mathrm{SI}],
\label{Ps_A3}
\end{eqnarray}
where $ {\bm \chi_{sa}}$ is calculated under the assumption of $\omega=\omega_r$. Through Equations (\ref{Pt})and (\ref{Ps_A2}), we have
\begin{equation}
\Sigma_s P^Q_s \simeq -\frac{4\pi \gamma}{\omega_r} = -2\gamma T,
\end{equation} 
where $T=2\pi/\omega_r$ is the wave period. $P^Q_s$ describes the energy transfer per unit of volume and per unit of wave energy in one mode period \citep{1998ApJ...500..978Q}. 

Comparing expressions (\ref{Ps_A1})$-$(\ref{Ps_A3}), we can see that expression (\ref{Ps_A3}) is valid under the condition $\gamma\ll \omega_r$. This implies that expression  (\ref{Ps_A3}) cannot describe the energy transfer associated with plasma waves with zero real frequency, for example, unstable waves driven by oblique firehose instability and ion/electron mirror instability \citep[e.g.,][]{2019ApJ...884...44S,2020ApJ...902...59S}. Although our expressions (e.g., expressions (\ref{Ps_1}), (\ref{Ps_2}), (\ref{Ps_A1}), (\ref{Ps_A2})) and Quataert's expression (\ref{Ps_A3}) have slightly different physical meanings, they are all helpful in quantifying the energy transfer between waves and particles.

\section{Typical parameters of proton beam instability}
Based on the data in Figures \ref{fig:r-vpb} and \ref{fig:r-vpb-Ps}, Table~\ref{table:parameters} summarizes characteristic values of the typical parameters in different proton beam instabilities. This table is helpful for understanding the evolution of proton beam instability in the inner heliosphere. For example, Subsection \ref{sub:6.1} discusses the evolution of both parallel and perpendicular plasma temperatures according to the energy transfer rates listed in Table~\ref{table:parameters}.



\clearpage

\end{document}